\documentclass[a4paper,UKenglish,cleveref, autoref, thm-restate]{amsart}

\usepackage{algorithmic}
\usepackage{amsmath, amssymb}
\usepackage{newfloat}
\usepackage{url}
\usepackage{multicol}
\usepackage{listings}
\usepackage{graphicx}
\usepackage{xcolor}
\usepackage{subcaption}

\usepackage{hyperref}
\usepackage{pgfplots}
\pgfplotsset{compat=1.18}
\usepackage{cancel}
\usepackage{amsthm}
\usepackage{listings}
\usepackage{tikz}
\usepackage{float}
\usepackage[english]{babel}
\newtheorem{lemma}{Lemma}

\usepackage{multirow}
\usetikzlibrary{trees,shapes,decorations}
\usetikzlibrary{positioning}
\usepackage[square,numbers,sort]{natbib}

\title{Timetable Nodes for Public Transport Network} 

\author{Andrii Rohovyi$^{1}$}
\thanks{$^{1}$Department of Computer Science and Engineering, University of New South Wales (UNSW), Sydney, Australia. \textit{a.rohovyi@unsw.edu.au}}
\author{Peter J. Stuckey$^{2,3}$}
\thanks{$^{2}$Department of Data Science and Artificial Intelligence, Monash University, Melbourne, Australia.}
\thanks{$^{3}$OPTIMA ARC Industrial Training and Transformation Centre, Melbourne, Australia. \textit{peter.stuckey@monash.edu}}
\author{Toby Walsh$^{4}$}
\thanks{$^{4}$Department of Computer Science and Engineering, University of New South Wales (UNSW), Sydney, Australia. \textit{t.walsh@unsw.edu.au}}

\newcommand{\pjs}[1]{\textcolor{blue}{\textsc{Pjs:} #1}} 
\newcommand{\ar}[1]{\textcolor{purple}{\textsc{AR:} #1}} 
\newcommand{\ignore}[1]{}

\pgfplotsset{compat=1.18}

\begin{document}

\begin{abstract}
Faster pathfinding in time-dependent transport networks is an important and challenging problem in navigation systems. There are two main types of transport networks: road networks for car driving \cite{katch, improving_time_dependent_ch} and public transport route network ~\cite{csa}. The solutions that work well in road networks, such as Time-dependent Contraction Hierarchies and other graph-based approaches, do not usually apply in transport networks  \cite{car_vs_transport}. In transport networks, non-graph solutions such as CSA ~\cite{csa} and RAPTOR \cite{raptor} show the best results compared to graph-based techniques. In our work, we propose a method that advances graph-based approaches by using different optimization techniques from computational geometry to speed up the search process in transport networks. We apply a new pre-computation step, which we call timetable nodes (TTN). Our inspiration comes from an iterative search problem in computational geometry. We implement two versions of the TTN: one uses a Combined Search Tree (TTN-CST), and the second uses Fractional Cascading (TTN-FC). Both of these approaches decrease the asymptotic complexity of reaching new nodes from  $\mathcal{O}(k\times \log|C|)$ to $\mathcal{O}(k + \log(k) + \log(|C|))$, where $k$ is the number of outgoing edges from a node and $|C|$ is the size of the timetable information (total outgoing edges). Our solution suits any other time-dependent networks and can be integrated into other pathfinding algorithms. Our experiments indicate that this pre-computation significantly enhances the performance on high-density graphs. This study showcases how leveraging computational geometry can enhance pathfinding in transport networks, enabling faster pathfinding in scenarios involving large numbers of outgoing edges.
\end{abstract}
\maketitle
\section{Introduction}

The transportation problem is a ubiquitous challenge in today's modern world, driving fierce competition within the market. Every company tries to achieve more efficient transport routing, which should be calculated quickly. The speed of response for pathfinding can be critical for business applications and plays a crucial role in competitiveness for routing services. In contrast to other fields in AI, such as Machine Learning and Deep Learning, development in this field has mostly lacked transparency. For instance, it is not easy to find open-source solutions in this area to compare results and iteratively improve them.

The transportation problem involves two main areas: the challenge of car travel on road networks and the issue of public transportation on transport graphs. It's interesting to note that completely different classes of algorithms are used to solve these two problems \cite{car_vs_transport}. Routing in road networks typically uses graph-based approaches like Time-dependent Contraction Hierarchy, whereas transit routing problems are usually addressed using non-graph approaches such as CSA \cite{csa} and RAPTOR \cite{raptor}.

\ignore{
Road networks have been extensively researched in terms of theoretical algorithms for planar graphs \cite{route_planning_in_transport_network}. While road networks are not strictly planar, they do have small separators. Therefore, it is likely that theoretically efficient algorithms for planar graphs will also perform well in road networks.
}

Road networks usually consist of roads with varying levels of importance. For example, when travelling long distances, we typically start with smaller arterial roads to get from home to the highway, then use the highway as the main route to our destination, and finally transition back to smaller roads \cite{car_vs_transport}. 

Due to this hierarchical structure \cite{car_vs_transport, route_planning_in_transport_network}, algorithms like Time-dependent Contraction Hierarchies (TCH) \cite{katch} or TRANSIT \cite{transit} are widely used in road networks. These approaches involve adding shortcuts to the graph, which enables quick responses to earliest arrival queries on large graphs.

In the meantime, when considering preprocessing steps for the public transport network, adding new shortcuts in the public transport graph might cause additional problems for the search process. Although these techniques can decrease the search space over nodes, they could also increase query time costs due to the higher complexity of the search over the edge ~\cite{contraction_hierarchy_transport_oliver}. This is because the graph becomes more dense. This problem happened with many other graph-based techniques \cite{car_vs_transport}. 

Recent research indicates that working directly with timetables like CSA \cite{csa} or with trips, such as RAPTOR \cite{raptor}, yields better performance in solving transport network problems. However, some researchers have shown that applying the Time-dependent Contraction Hierarchy to the long-distance train connections problem with timetable data still shows promising results \cite{contraction_of_timetable, route_planning_in_transport_network}.

In this paper, we aim to advance the use of graph-based methods for solving transport routing problems by implementing speed-up techniques from computational geometry. We introduce a new technique for handling high-density graphs. We develop a method called timetable nodes (TTN) which enables the application of binary search at the node level instead of the edge level. Two versions of TTN are created: Combined Search Tree (TTN-CST) and Fractional Cascading (TTN-FC). Moreover, Fractional Cascading (TTN-FC) has also been adapted to the search technique over the Contraction Hierarchies Graph. This adaptation significantly improves query runtime.

\section{Preliminaries}
This section establishes basic terminology and introduces foundational algorithms
\subsection{Terminology}

\subsubsection*{Transport Network}
A Transport Network is a complex system of interconnected infrastructure that facilitates people's movement from one location to another. It provides essential connections between various places, enabling efficient and effective transportation and communication. Instead of classical graphs, the edges in the transport network are modelled as functions of time $f(t)$.

In its mathematical formulation, this graph can be presented as $G=(V, E, F, T)$, where $V$ is the set of nodes, $E$ is the set of edges, where  $E \subseteq V \times V$, $f \in F$ is a set of functions, where for edge $e = (v,v')$, given start time $t$ at node $v$, $f_e(t)$ returns the arrival time at $v'$ using edge $e$. \ignore{Each function could be activated by a specific start time $t \in T$.}

\subsubsection*{Multimodal Transport Network}

One type of Transport Network is the Multimodal Transport Network. It is a system where each function $f(t)$ represents different modes of transportation, such as cars, buses, trains, walking, etc.

In our study, we focused on a system that depicted transportation modes for public transport and walking.

There are various ways in which $f \in F$ could be implemented in the Multimodal Transport Network. Our research employs its representation as an Arrival Time Function~\cite{contraction_hierarchy_transport_oliver}.

\subsubsection*{Arrival Time Functions (ATF)}
ATF ~\cite{contraction_hierarchy_transport_oliver} are a way of representing the travel time functions $f(t)$ in a Multimodal Transport Network. The arrival time function provides information about the expected or actual arrival (and implicitly departure) times of transportation services at specific nodes in the transport network. 

These functions have the next set of features:
\begin{itemize}
    \item $\forall f \in F, \forall x, y \in T, x \geq y \Rightarrow f(x) \geq f(y) $
    \item $\forall f \in F, \forall t \in T, f(t) \geq t $
    \item $\forall f, g \in F, f \circ g \in F $
    \item $\forall f, g \in F, \min ( f, g ) \in F $
\end{itemize}

\ignore{Unlike other techniques such as ULTRA~\cite{ultra}, we do not need to insert extra nodes to cover walking connections. Information about both walking and bus profiles is combined into a single function.
\pjs{I found this too vague, perhaps have a forward pointer to an example diagram?} 
\ar{I ommit this part to reduce the size of paper to 12 pages}
}

\subsubsection*{Walk profile}
The walk profile function calculates the arrival time for pedestrians trying to get from one node (e.g., bus stop, train station) to another node within the network by foot. This information is essential for passengers who need to transfer between different transportation services or modes of transport during their journey. The classical walk profile function looks the following way:
$$f_{v_1,v_2}(t) = t + w(v_1,v_2)$$
where $w(v_1,v_2)$ is the walk duration to walk from node $v_1$ to node $v_2$. Note that we may disallow walking along edges which are too long, by setting $w(v_1,v_2) = \infty$.

\subsubsection*{Timetable}

The timetable is the set of departure and arrival times from one stop to another, where the arrival time is greater than or equal to the departure time. $$C = \{(d, a)| d, a \in T, a \geq d\} $$ 

\subsubsection*{Bus profile}
 The bus profile function calculates the arrival time for pedestrians trying to get from one node (e.g., bus stop, train station) to another node within the network by bus, train, or any other public transport that uses timetables. The classical bus profile function looks as follows:
$$f_{v_1,v_2}(t) = min\{a|\exists(d, a)\in C_{v_1,v_2}, d \geq t\},$$
where $C_{v_1,v_2}$ 
are the timetable connections along edge $(v_1,v_2)$, 
given as pairs $(d,a)$ of departure time $d$ and arrival time $a$.

\subsubsection*{Size of ATF}

\begin{figure}[h]
\begin{multicols}{2}
    \begin{tikzpicture}[scale=1]
    \begin{scope}[every node/.style={circle,thick,draw,minimum size=15pt,inner sep=2pt}]
        \node (A) at (-1, 0.5) {A};
        \node (B) at (3, 2) {B};
        \node (D) at (3, -1.5) {D};
        \node (C) at (3, 0.5) {C};
    \end{scope}
    \begin{scope}[every node/.style={fill=white,circle},
                  every edge/.style={draw=black, thick}]
        \path [->] (A) edge (B);
        \path [->] (A) edge (C);
        \path [->] (A) edge (D);
    \end{scope}
    \end{tikzpicture}

    \columnbreak

    \footnotesize
    \textbf{A $\rightarrow$ B:}
    \begin{itemize}
        \item Walking time: 40 minutes
        \item Timetable:  (14:00, 14:20), (15:15, 15:20)
    \end{itemize}

    \textbf{A $\rightarrow$ C:}
    \begin{itemize}
        \item Walking duration: $\infty$ minutes
        \item Timetable: (13:30, 13:50), (18:00, 18: 20), (20:10, 20: 50)
    \end{itemize}

    \textbf{A $\rightarrow$ D:}
    \begin{itemize}
        \item Walking duration: 20 minutes
        \item Timetable: (12:00, 12:30), (12:45, 13:30), (15:15, 15:30), (16:05, 16:30)
    \end{itemize}
    \normalsize

\end{multicols}
\caption{\label{fig:ex}A node $A$ in a transport network, with outgoing edges to $B$, $C$, and $D$.}
\end{figure}
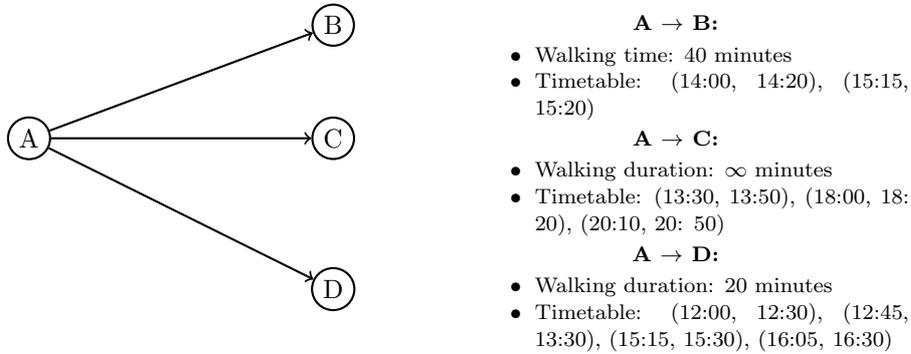

The size of the ATF functions is determined by the number of elements in the timetable.

Let's take a look at the example graph segment shown in Figure~\ref{fig:ex}. In this graph, the node $A$ has $3$ outgoing arcs. The edge $A \rightarrow B$ is denoted by a walking connection with a 40-minute duration and a bus with a timetable size of two $size(f_{A, B})=size(C_{A, B})=2$. Similarly, the edge $A \rightarrow C$ is represented by the timetable size of three, but it is not feasible to reach it by walking, as the walking duration is set to infinity $size(f_{A, C})=size(C_{A, C})=3$. The edge $A \rightarrow D$ is depicted by the size of timetable four, along with a walking connection of 20 minutes $size(f_{A, D})=size(C_{A, D})=4$.

\subsubsection*{ATF evaluation}

To evaluate the Arrival Time Function, we employ a binary search over the timetable with bus profiles and then compare the resulting arrival time with the walking profile. The complexity of this operation is $O(\log(\text{size}(f_{v_1,v_2})))$, where $\text{size}(f_{v_1,v_2})$ represents the size of the timetable for this edge.
$$f_{v_1,v_2}(t) = min(t + w(v_1,v_2), 
                   min(a|\exists(d, a)\in C_{v_1,v_2}, d \geq t))$$

Let's analyze the asymptotic complexity of such evaluation:
Assuming  the search over timetable $C_{v_1,v_2}$ is binary, the complexity of evaluation is we have the next 
$\mathcal{O}(min\{1, log(|C|)\}) = \mathcal{O}(log(|C_{v_1,v_2}|))$.

\subsubsection*{Chaining}
Chaining in the transport network is an operation of computation of the new function, which covers a path over two edges. For instance, we have an edge from node $u$ to node $v$, which is represented as a function $f_{uv}$ and an edge from node $v$ to node $w$, which is defined as a function $f_{vw}$, then chaining is a new function from node $u$ to $w$, which is equal to $f_{uw}(t) = f_{vw} \circ f_{uv}(t)=f_{vw}(f_{uv}(t))$.

\subsubsection*{Merging}
Merging in a Transport Network minimises two functions $f_{uv}(t)$ and $f'_{uv}(t)$ over the same edge $uv$. In this case, we have two parallel edges and need to build a new function $f''_{uv}(t) = min(f_{uv}(t), f'_{uv}(t))$, which will preserve the shortest path traversal across this edge.

\subsubsection*{Path}

A path $P$ in a transport graph between node $s$ and node $d$ is a sequence of nodes $\langle v_0, v_1, ... , v_k \rangle$, where $ k \in \mathbb{N}^+$ , $v_0 = s$ 
and $v_k = d$ and $\forall i \in \{0, 1, 2... ,k-1\}  \exists e_{v_i, v_{i+ 1}} \in E, \exists v_i \in V, \exists v_{i+1} \in V$. In the transport network the length or cost for path
$|P| = f_{P}(t)$, where $ f_{v_0,v_k}(t)=f_{v_{k-1},v_k}(f_{v_{k-2},v_{k-1}}(...f_{v_1,v_2}(f_{v_0,v_1}(t))...))$ 
is dependent on start time $t \in T$.

Let $sp(s, d, t)$  denote the shortest path from node $s$ to node $d$, when we start our movement from node $s$ at the time $t \in T$.

\subsubsection*{Merged timetable}
A merged timetable is another way of representing the public transport system and is used in non-Dijkstra algorithms, such as the Connection Scan Algorithm (CSA). The merged timetable ($C_m$) is a tuple of $(V, C, W, Tr)$, where $V$ represents the set of stops (similar to the set of nodes in the previous representation), $C$ represents the timetable between two stops, $W$ represents footpath connections between stops, and $Tr$ represents scheduled transport between stops.

The size of the merged timetable is the number of elements in $C_m$.

\subsection{Algorithms}

\subsubsection*{Connection Scan Algorithm (CSA)}

The Connection Scan Algorithm (CSA) \cite{csa} efficiently answers queries for timetable information systems. It is a non-graph approach that is specialized for timetable systems. We used it for a comparison analysis. The algorithm puts all information into one merged timetable $C_{m}$, runs a binary search once over the whole timetable to find the closest connection, and then iteratively checks if the next connections could be a part of our journey. It stores the arrival time for every station and stops when we explore a connection with a departure time greater than the arrival time at the target station.

\subsubsection*{Dijkstra (Dij.)}
Dijkstra's algorithm \cite{dijkstra1959note} is a graph-based pathfinding algorithm used to find the shortest path between two nodes in a weighted graph. The time complexity of Dijkstra's algorithm is $\mathcal{O}(|E| + |V| \times \log(|V|))$ \cite{fibbonaci}.

\subsubsection*{Contraction Hierarchies (CH)}
\sloppy

Contraction Hierarchies is a graph augmentation method that helps speed up future pathfinding \cite{contraction_hierarchy}. 
The idea of this algorithm is to iteratively contract nodes from the graph and insert new edges that represent shortest paths through the contracted node. In broad strokes:

\begin{enumerate}
  \item Set up hierarchical order $\mathcal{L}$ of the nodes contraction. While preprocessing and query algorithms work for any contraction ordering, it is preferable to use one that produces as few shortcuts as possible. This leads to a small space requirement and faster query time. The problem of minimizing the number of edges of a given graph’s Contraction Hierarchies is APX-hard, so it is computed using heuristics \cite{milosavljevic2012optimal}. In our research, we used linear combinations of the following factors as heuristics to determine the order of contraction:
  \begin{enumerate}
  \item Edge difference. The immediate cost of the number of edges is calculated as the number of shortcuts inserted minus the number of edges removed.
  \item Depth. When starting off, the depth of a vertex $u$ is 0. When a vertex $u$ is contracted, the depth of its neighbour $v$ is updated to be the maximum of its current depth and the depth of $u$ plus 1. This updated depth value serves as an upper limit for the length of a path when making queries, and it helps in choosing vertices more evenly.
  \end{enumerate}
  \item Iteratively select the lowest node $v$ in the hierarchy $\mathcal{L}$, which hasn't been previously contracted.
  \item Contract selected node $v$, adding a new shortcut edges $e_{uw}$, which will cover paths from each in-neighbor $u$ to each out-neighbor $w$ passing through $v$. 
  Obviously, $v$ is lower in the  hierarchy than $u$ or $w$.
  \item After contracting all nodes, in order to limit the number of shortcuts added, we perform a shortest path search to determine if a shortcut may be unnecessary. This procedure is called witness search, and it is used to exclude shortcuts that do not represent the optimal path.
\end{enumerate}

Every contracted graph has two types of edges: 
\begin{itemize}
  \item Upward edge $e_{uw}\uparrow$, if node $w$ lexically larger then node $u$, $\mathcal{L}_u < \mathcal{L}_w$
  \item Downward edge $e_{uw}\downarrow$, if node $u$ lexically large then node $w$, $\mathcal{L}_u > \mathcal{L}_w$
\end{itemize}

\subsubsection*{Time-dependent Contraction Hierarchies (TCH)}

TCH \cite{katch} is an adaptation of CH for the transport network, where every edge is represented as a function $f(t)$ over the time $t \in T$. In our TCH implementation, we omitted a witness search due to its computational expense on the graph with the ATF functions we use.

In CH and TCH, a new shortcut edge $e_{uw}$ is added if both nodes $u$ and $w$ appear later in the hierarchy than the intermediate node $v$, and $e_{uw}$ is optimal and equivalent to the path $\langle u;v;w \rangle$. Therefore, for every pair of edges $(e_{uv};e_{vw})$, there must exist a cost-equivalent upward edge $e_{uw}\uparrow$ (i.e., $u < \mathcal{L}_w$) or downward edge $e_{uw}\downarrow$ (i.e., $u > \mathcal{L}_w$) if $\langle u;v;w \rangle$ is a shortest path from $u$ to $w$ for some $t \in T$. 

The following results,  paraphrased here,  are due to \cite{contraction_hierarchy}:

\begin{lemma}[TCH-path]
    For every optimal path $\textit{sp}(s;d;t)$ in $G$, there is a cost equivalent TCH-path $\langle s;\ldots;k \ldots;d \rangle$ whose prefix $\langle s;\ldots;k \rangle$ is an up path (i.e., $s < \mathcal{L}_k < \ldots < \mathcal{L}_{k+1} < \ldots < L_d$), and the suffix $\langle k;\ldots;d \rangle$ is a down path (i.e., $\mathcal{L}_s > \ldots > \mathcal{L}_{s+1} > \ldots > \mathcal{L}_k > \ldots > \mathcal{L}_{k+1} > \ldots > \mathcal{L}_d$).
    \qed
\end{lemma}

\textit{Corollary 1. (apex node):} Every TCH-path contains a node $k$ that is lexically the largest node in the hierarchy among all nodes on the TCH-path. \qed

Contracting does not affect the optimal shortest path. The fundamental concept behind the TCH is to utilize shortcut edges that can directly bypass one or more intermediate nodes, leading to faster pathfinding.

\subsubsection*{Forward search (FS)}

FS \cite{forward_search} is a pathfinding algorithm using forward (A* \cite{hart1968astar}) search over the contraction hierarchy graph, which utilizes CH-graph characteristics to optimize query runtimes.  Usually search in CH-graphs uses bi-directional search~\cite{katch}, the use of forward search means we can use heuristics to reduce the search.

\ignore{
Consider a scenario with a given source node $s$ and destination node $d$.  During the expansion of a search node $n$ with its predecessor $p(n)$, the search algorithm on the Contraction Hierarchies typically generates successors $s(n)$ that can be categorized into the following types:
\begin{enumerate}
    \item Up-Up successors: $p(n) < L_n$ and $n < L_s(n)$;
    \item Up-Down successors: $p(n) < L_n$ and $n > L_s(n)$;
    \item Down-Down successors: $p(n) > L_n$ and $n > L_s(n)$;
    \item Down-Up successors: $p(n) > L_n$ and $n < L_s(n)$.
\end{enumerate}
The underlying concept reveals that an optimal path can comprise the Up-Up, Up-Down, and Down-Down successors. However, including a Down-Up successor would violate the Contraction Hierarchies (CH) path property, as directional changes after the apex node are not permitted. Consequently, to maintain the CH-path property, the authors in \cite{forward_search} introduced a modified search approach that only considers paths adhering to a simple pruning rule called UTD (Up-Then-Down).

It is essential to note that an optimal path is always an Up-Down path (as implied by Lemma 1), and it features an apex node that is lexically larger than all the other nodes on the path.
}
For Forward Search we can precompute down-reachable bounding boxes for every node $v$ that are large enough to contain all the nodes reachable from $v$ by only down edges. 
During the search, before taking a down edge from $v$ 
we can check if the target node is contained in the down-reachable bounding box of $v$, if not we do not need to consider taking any down edge from $v$.
These pruning rules significantly improve the search process~\cite{forward_search,improving_time_dependent_ch}.
\ignore{
Additional improvements that we can add to the Forward search algorithm include precompiling in advance bounding boxes for each node $v$. These containers collect information about the set of all nodes, which are reachable during down movement from node $v$. These containers could be calculated with a depth first search algorithm in the preprocessing stage. These heuristics significantly improve the search process~\cite{forward_search,improving_time_dependent_ch}.}
\section{Timetable nodes (TTN)}
To speed up the edge evaluation process, we modify the pathfinding algorithm and move the evaluation process to the node level \cite{multi_dijkstra}. The idea behind this is to store information about departure time not on the edge level, as is the case in a graph with timetable edges (TTE), but on the node level and, during the pathfinding process, evaluate this level.

This idea is inspired by an iterative search problem in computational geometry \cite{combined_search_tree}. This problem could be formulated in the following way. Let's assume that we have $k$ catalogues and query value $x$, and we try to simultaneously to find the least value larger than or equal value to $x$ in all $k$ catalogues. This task appears in computational geometry problems \cite{fractional_cascading}, such as:
\begin{itemize}
  \item Given a collection of intervals on the line, how many of them intersect  an arbitrary query interval?
  \item Given a polygon $P$, which sides of $P$ intersect an arbitrary query line?
  \item Given a collection of rectangles, which of them contains an arbitrary query point? 
\end{itemize}

\subsection{Combined Search Tree}

One of the approaches in resolving iterative search problems is called the Combined Search Tree~\cite{combined_search_tree}. 
The concept behind this method is to combine all departure times from schedules for edges leaving node $v$. This allows us to find connections by performing a binary search on this combined list of departure times. After this step, we can refer to the precomputed matrix $k$ times. This matrix contains data about the connections between departure times in the combined list and the departure times in the outgoing edges, with $k$ representing the number of outgoing edges.

This idea finds good application in search over the transport network graph. Recall the graph segment shown in Figure~\ref{fig:ex} that we have the graph with nodes $A$, $B$, $C$
and $D$ and edges between nodes $A$ and nodes $B$, $C$ and $D$. 

\ignore{
Edge $A \rightarrow B$ contains timetable information about the bus, which departs at 13:00 from node $A$ and arrives at 13:10 at node $B$. Another bus goes at 15:00 from node $A$ and arrives at 15:00 at node $D$. Edge $A \rightarrow C$ contains timetable information about the one bus with departure at 12:00 and arrival at 12:10 and the second one with departure at 14:00 and arrival at 14:10. In this case, node $A$ will contain a concatenated timetable with information about the whole edges, which is outgoing from it.

\begin{center}
\begin{tikzpicture}[scale=1]

\begin{scope}[every node/.style={circle,thick,draw}]
    \node (A) at (-2, 5) {A};
    \node (B) at (2.5, 8) {B};
    \node (C) at (2.5, 2) {C};
    \node (D) at (2.5, 5) {D};
\end{scope}

\begin{scope}[every node/.style={fill=white,circle},
              every edge/.style={draw=black, thick}]
    \path [->] (A) edge node[align=center] {$(\text{12:00}, \text{12:10})$ \\ $(\text{14:00}, \text{14:10})$} (C);
    \path [->] (A) edge node[align=center] {$(\text{13:00}, \text{13:10})$} (B);
    \path [->] (A) edge node[align=center] {$(\text{15:00}, \text{15:10})$} (D);
\end{scope}

\node[left=0.01cm of A, align=center] (timetableA) {
  TimetableA:  \\
    $(\text{12:00}, \text{12:10})$ \\
    $(\text{13:00}, \text{13:10})$ \\
    $(\text{14:00}, \text{14:10})$ \\
    $(\text{15:00}, \text{15:10})$
};

\end{tikzpicture}
\end{center}
}
We can created the sorted concatenation of departures connections from node $A$ as: 12:00, 12:45, 13:30, 14:00, 15:15, 16:05, 18:00, 20:10.

In a pre-computation step, we calculate a matrix with information about the connection between the node and edges departure times. For the graph segment of Figure~\ref{fig:ex}, it will be computed as shown in Table~\ref{tab:ex}, where $-$ indicates there is no possible connection available.

\begin{table*}[t]
\caption{Combined Search Tree matrix for node $A$ of Figure~\ref{fig:ex}.\label{tab:ex}}
\tiny
\begin{tabular}{p{1.2cm}|p{1cm}p{1cm}p{1cm}p{1cm}p{1cm}p{1cm}p{1cm}p{1cm}}
    \multicolumn{1}{c}{} & \multicolumn{3}{c}{} \\
    Node A & 12:00 &  12:45 & 13:30 & 14:00 & 15:15 & 16:05 & 18:00 & 20:10 \\
    \hline
    $A \rightarrow B$ & 14:00 & 14:00 & 14:00 & 14:00 & 15:15&---&---&--- \\
    $A \rightarrow C$ &13:30& 13:30& 13:30 & 18:00&18:00&18:00&18:00&20:10 \\
    $A \rightarrow D$ & 12:00   & 12:45 & 15:15  &  15:15 &15:15&16:05&---& ---\\
\end{tabular}
\end{table*}

 This matrix helps us to match the result of a binary search over a timetable node with departure time for some exact edge. For instance, if our agent arrives at node $A$ at 13:15, the closest bus at the node timetable will arrive at 13:30 (third column of matrix). After just making 3 look-ups, he was able to find the closest bus to node $B$  at 14:00 (first row), node $C$ at 13:30 (second row), and node $D$ at 15:15 (third row). 

A simple explanation of this algorithm follows. When you stay at one stop and search for the next buses to each  reachable stop, it will be much faster to look at one big timetable and find these buses at once than to look at separate timetables for each arrival stop.

The complexity of such a procedure for node $v$ is $\mathcal{O}(k+log(k\times|C|)$) = $\mathcal{O}(k+\log(k) + \log(|C|)$)  where $k$ is the number of outgoing edges from $v$, and $|C|$ is the number of outgoing connections on each edge from $v$.  We make a single binary search on the array of connection times, and then lookup the $k$ earliest connections. 

In contrast if we store separate departures for each outgoing edge the complexity is 
$O(k \times \log(|C|)$ since we must separately look up the $k$ separate timetables each of size $|C|$.

Of course, there is a cost to maintaining the combined search tree, as it requires storing a matrix that corresponds departure times over the node with departure times in edges.
This requires additional space in $\mathcal{O} (k^2 \times |C| \times n)$, where $n$ is the number of nodes in the graph,
since for each node in a graph, we will need to precompute a matrix of size $\mathcal{O}(k \times k * |C|)$.

\subsection{Fractional Cascading}

We now introduce an advanced technique from computational geometry called Fractional Cascading~\cite{fractional_cascading} which allows us to
find the next connection for all outgoing edges with just $\mathcal{O} (k \times |C| \times n)$ of additional space 
but retaining the same time complexity $\mathcal{O}(k + \log(k) + \log(|C|))$ as the combined search tree. 

Given a series of ordered lists $D_0, D_1, \ldots, D_k$ where each $D_i$ is a list of elements sorted in increasing order, the goal of Fractional Cascading is to preprocess these lists into a tree structure that allows for efficient search queries. A search query involves finding the position of a given value $t$ in each of the lists $D_0, D_1, \ldots, D_k$.

The preprocessing should enable search operations such that if we find the position of $t$ in $D_i$ in time $\mathcal{O}(\log size(D_i))$, we can find the positions of $t$ in $D_{i+1}, D_{i+2}, \ldots, D_k$ in total time $\mathcal{O}(k)$, where $k$ is the number of subsequent lists.

The preprocessing involves the following steps:

\begin{enumerate}
 \item Initiation: Set up augmented list $M_k$ equal to original list $D_k$, where $k$ is amount of lists.
 \item Lists Augmentations: For each list $D_i$ create an augmented list $M_i$ by adding a fraction of elements from the subsequent list $M_{i+1}$. Specifically, for each element in $C_i$, merge it with every second element in $M_{i+1}$.
 
 \item Bridges Setup: 
 \begin{itemize}
     \item
     For each element $d_{i,j}^m$ in the augmented list $M_i$, we build a linker, which is called a "bridge" to the element $d_{i+1,k}^m$ in $M_{i+1}$, where $d_{i+1,k}^m = argmin_{d_{i+1,k}^m \in M_{i+1}} |d_{i,j}^m - d_{i+1,k}^m|$. This bridge enables us with constant time complexity $\mathcal{O}(1)$ to locate the closest position to the element in the upcoming augmented list $M_{i+1}$, after identifying the closest position in list $M_i$.
     \item 
     In the same way, for each element $d_{i,j}^m$ in the augmented list $M_i$, we build a bridge to the element $d_{i, k}^c$ in the original list $D_i$ in such way that $k = argmin_{d_{i,k}^c \in C_i} |d_{i,j}^m - d_{i,k}^c|$. This bridge allows us to extract location of original departure time during the search process.
 \end{itemize}

 \item Search Procedure: To search for $x$ in the original lists $D_0, D_1, \ldots, D_k$:
 . \begin{itemize}
   \item Start by searching for $x$ in $M_0$ using binary search. Use the pointer to find a position in the original list $D_0$
    \item Use the pointers to find the position of $x$ in $M_1$, then $M_2$, and so on, without performing a full binary search in these lists. Consequently, use the same pointers to find positions in the original lists $D_{0}, D_{1}, \ldots, D_k$.
   \end{itemize}
\end{enumerate}

For instance, in the case we present at Figure 1, the Fractional Cascading structure of the information will look in the following way:

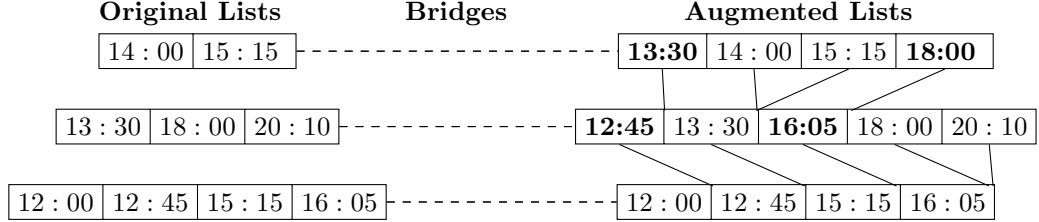
\begin{tikzpicture}[
  level distance=3cm,
  sibling distance=5cm,
  edge from parent/.style={draw, -latex},
  every node/.style={rectangle split, rectangle split horizontal, align=center, draw, rectangle split draw splits=true}
]

\node[draw=none] at (0.5,0.5) {\textbf{Original Lists}};
\node[draw=none]  at (4,0.5) {\textbf{Bridges}};
\node[draw=none]  at (8.5, 0.5) {\textbf{Augmented Lists}};

\node[rectangle split parts=2] (A) at (0,0) {  $14:00$ \nodepart{two} $15:15$ };
\node[rectangle split parts=3] (B) at (0,-1) { $13:30$  \nodepart{two} $18:00$ \nodepart{three} $20:10$};
\node[rectangle split parts=4] (C) at (0,-2) {$12:00$ \nodepart{two} $12:45$ \nodepart{three} $15:15$ \nodepart{four} $16:05$};

\node[rectangle split parts=4] (A2) at (8,0) { $\textbf{13:30}$ \nodepart{two} $14:00$ \nodepart{three} $15:15$ \nodepart{four} $\textbf{18:00}$ };
\node[rectangle split parts=5] (B2) at (8,-1) {$\textbf{12:45}$ \nodepart{two} $13:30$ \nodepart{three} $\textbf{16:05}$  \nodepart{four} $18:00$ \nodepart{five} $20:10$};
\node[rectangle split parts=4] (C2) at (8,-2) {$12:00$ \nodepart{two} $12:45$ \nodepart{three} $15:15$ \nodepart{four} $16:05$};

\draw (A2.one south) -- ++(0.03,-0.55);
\draw (A2.four south) -- ++(-1.23,-0.55);
\draw (B2.one south) -- ++(1.23,-0.55);
\draw (B2.three south) -- ++(1.23,-0.55);

\draw (A2.two south)  -- ++(0.04,-0.55);
\draw (A2.three south) -- ++(-1.21,-0.55);
\draw (B2.two south) -- ++(1.23,-0.55);
\draw (B2.four south) -- ++(1.24,-0.55);
\draw (B2.five south) -- ++(0.05,-0.55);

\draw[dashed] (A.two east) -- (A2.two west);
\draw[dashed] (A.five east) -- (A2.five west);
\draw[dashed] (B.two east) -- (B2.two west);
\draw[dashed] (B.four east) -- (B2.four west);
\draw[dashed] (C.two east) -- (C2.two west);
\draw[dashed] (C.four east) -- (C2.four west);

\end{tikzpicture}
\sloppy

In the example above, the bottom list $M_2$ is equal to the list of departure times over the edge $A \rightarrow D$, the previous list $M_1$, which is equal to $12:45, 13:30, 16:05, 18:00, 20:10$ is equivalent to the merge of the list of departure times in edge $A \rightarrow C$ with each second element in list $M_2$. And the top list $M_0$ is equal to the merge of $13:30, 18:00$ 
with the list of departure times in 
edge $A \rightarrow B$.

We can look up the next services for each edge from $A$ when we arrive there at time 13:15 by binary search in the first list to discover the 13:30, the first service to B is the next entry 14:00. We then follow the pointer from 13:30 to the list below, and find the next entry to C is 13:30, we then follow the pointer from 13:30 to the augmented list D. We see the following entry in the augmented list D is 12:45. Still, since our arrival time is 13:15, we look at the next entry, which is 15:15 and then find the following service to D is at 15:15.

In this case, the top list $M_0$ consists of 4 elements, which is less than the size of the merged list of original department times (8 elements). In the general case, the size of $M_i$ is equal to or less than:
$$ |D_{v,v_i}| + \frac{1}{2} \times |D_{v,v_{i+1}}| + ... + \frac{1}{2^j} \times |D_{v,v_{i+j}}|+...,$$
where $|D_{v,v_i}|$ is the number of the departure times on outgoing edge $(v,v_i)$ from node $v$. $|D_{v,v_i}|$ is equal to the size of timetable $|C_{v,v_i}|$.

Based on this formula, the maximum augmented list $M_0$ could be as large as the combined list in the worst-case scenario. However, in many cases, it could be even smaller. The asymptotic complexity of the binary search in $M_0$ is $\mathcal{O}(\log(k\times|C|)) = \mathcal{O}(\log(k) + \log(|C|))$, where $k$ represents the number of outgoing edges and $|C|$ represents the average size of timetables. The total complexity of the node evaluation process is the sum of the complexity of the binary search in $M_0$ and the complexity of finding the closest departure time in lists $ M_1, \ldots, M_k$ which equals $\mathcal{O}(k + \log(k) + \log(|C|))$.

The total size of the data structure is equal to the following:
$$ \sum_{i=1}^{k} M_i = \sum_{i=1}^{k} |D_i| (1 + \frac{1}{2} + \frac{1}{4}+...) \leq 2 \times k \times |D| = 2 \times k \times |C|,$$  
where $|C_i|$ is the size of the outgoing edge $i$ from node, $|D_i|$ is the number of departure times in timetable $C_i$ and $k$ is average amount of outgoing edges. Therefore, the total memory usage of the mentioned technique will be in $n$ times bigger and equal to $\mathcal{O}(k \times |C| \times n)$, where $n$ is the total amount of nodes in the graph.

In each iteration of Fractional Cascading, we verify if the bus profile arrives before the walking profile, provided that the walking profile exists. If there are walking connections without a timetable connection, we address them using a traditional method by systematically searching over them after applying Fractional Cascading.

\subsubsection{Sorting strategies}
Different sorting strategies are used for setting up lists in Fractional Cascading. The main ones involve sorting lists in ascending order (ASC) and placing smaller lists at the bottom of the fractional cascading hierarchy or sorting in descending order (DSC) and placing larger lists at the bottom of the fractional cascading hierarchy.

\subparagraph{Contraction Hierarchical Sorting}

We have developed a special adaptation of the Timetable Nodes using the Fractional Cascading (TTN FC) technique for the Forward Search algorithm called Contraction Hierarchical Sorting (CHHIER). We reverse the sequence of Fractional Cascading trees to match the hierarchy of nodes connected through edges in the TCH graph. This means that the list of departure times of the outgoing edges, which is connected to the nodes with the highest hierarchy, will be located at the bottom of the Fractional Cascading tree. This approach enables us to streamline the Forward Search while moving downward. For example, when the augmented list is linked to the edge connected to the node higher in the hierarchy than the current node, we can end the extraction of departure times over the Fractional Cascading tree.

For instance, in Figure 1, if node $A$ has a hierarchy of 54, node $B$ has a hierarchy of 60, node $C$ has a hierarchy of 52, and node $D$ has a hierarchy of 15, the Fractional Cascading tree will look as follows:

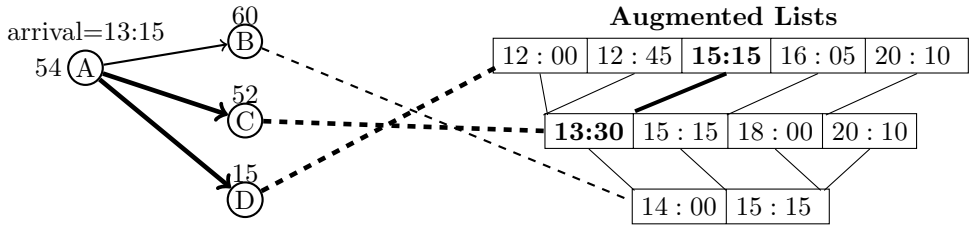
\begin{figure}[h]
\centering
\begin{multicols}{2}
    \begin{tikzpicture}[scale=0.7]
    \begin{scope}[every node/.style={circle,thick,draw,minimum size=10pt,inner sep=1pt}]
        \node (A) at (-1, 1.5) {A};
        \node (B) at (2, 2) {B};
        \node (D) at (2, -1) {D};
        \node (C) at (2, 0.5) {C};
    \end{scope}
    \begin{scope}[every node/.style={fill=white,circle},
              every edge/.style={draw=black, thick}]
    \path [->] (A) edge (B);
    \path [->] (A) edge[draw=black, line width=1.75pt] (C);
    \path [->, line width=2pt] (A) edge[draw=black, line width=1.75pt] (D); 
\end{scope}
    \begin{scope}[every edge/.style={draw=black, thick, dashed}]
        \path [-] (B) edge (9.2,-1);
        \path [-] (C) edge[draw=black, line width=1.75pt] (7.7, 0.3);
        \path [-] (D) edge[draw=black, line width=1.75pt] (6.7, 1.5);
    \end{scope}
    \node at (-1.7, 1.5) {54};
    \node at (-1, 2.2) {arrival=13:15};
    \node at (2.0, 1.0) {52};
    \node at (2.0, 2.5) {60};
    \node at (2.0, -0.5) {15};
    \end{tikzpicture}

    \columnbreak

    \begin{tikzpicture}[
      level distance=2cm,
      sibling distance=3.5cm,
      edge from parent/.style={draw, -latex},
      every node/.style={rectangle split, rectangle split horizontal, align=center, draw, rectangle split draw splits=true}
    ]

    \node[draw=none]  at (6.5, 0.5) {\textbf{Augmented Lists}};

    \node[rectangle split parts=5] (A2) at (6,0) { $12:00$ \nodepart{two} $12:45$ \nodepart{three} $\textbf{15:15}$ \nodepart{four} $16:05$ \nodepart{five} $20:10$ };
    \node[rectangle split parts=4] (B2) at (6,-1) {$\textbf{13:30}$ \nodepart{two} $15:15$ \nodepart{three} $18:00$  \nodepart{four} $20:10$};
    \node[rectangle split parts=2] (C2) at (6,-2) {$14:00$ \nodepart{two} $15:15$ };

    \draw (A2.one south) -- ++(0.1,-0.55);
    \draw (A2.two south) -- ++(-1.2,-0.55);
    \draw[ultra thick, line width=1.5pt] (A2.three south) -- ++(-1.2,-0.5);
    \draw (A2.four south) -- ++(-1.2,-0.55);
    \draw (A2.five south) -- ++(-1.2,-0.55);
    \draw (B2.one south) -- ++(0.6,-0.55);
    \draw (B2.two south) -- ++(0.6,-0.55);
    \draw (B2.three south) -- ++(0.6,-0.55);
    \draw (B2.four south) -- ++(-0.6,-0.55);

    \end{tikzpicture}

\end{multicols}
\end{figure}

In this scenario, departure times for the routes A to B are listed at the bottom level of the FC tree, as node B is lexically larger than other nodes connected by outgoing edges from node A. Departures from A to C are listed next. In contrast, at the top level, we list the departure times for routes A to D, as node D is lexically smaller than C and D. Consequently, when we run the Forward Search algorithm and the search goes downward, we could stop it without exploring the edges A to B, as node B lexically larger than node A.

For example, if the arrival time to node A was 13:15, by binary searching in the first augmented list, we found the closest departure time of 15:15 and, consequently, 15:15 in the original list after using the bridge. We then follow the pointer 15:15 to the list below and extract the departure time 13:30. At this point, we truncate the search in the FC tree, as we do not need to find the departure time in the bottom list since it is related to the edge A to B.

In case of upward movement, we do not need to truncate it and could extract the next departure time, 14:00, from the bottom list. This sorting method helps us speed up the search over the FC graph.

We implemented this modification in the Forward Search algorithm utilized in our experiment. However, it could be incorporated into any other pathfinding algorithm over the CH-graph that utilizes node hierarchy during the search, such as Bidirectional Dijkstra.

\section{Experiments}

All calculations and research have been done using Python on the MacBook Pro with Apple M3 Max chip, which has a 16-core CPU, 40-core GPU, and 128GB Unified Memory.  For reproducibility, all of our implementations are available \footnote{\url{https://github.com/andrii-rohovyi/timetable_nodes.git}}.

\subsection{Real city analysis}
\ignore{
\pjs{Shouldnt we also compare with CSA?} 
\ar{I think, it is a good idea. Big thank you for the suggestion. Should I rewrite CSA on Python for this paper, as all research has been done on it? Or does it make sense to add this comparison in future work after rewriting everything on C++ and compare with the original implementation ?}
\pjs{Well you could try a Python CSA implementation, but it might be criticised for not necessarily having all the benefit of a language with more memory control, still I think its worth adding even in Python?} 
}

Our research used a public dataset of about 25 cities published by Kujala~\emph{et al}~\cite{dataset}. The complete list of cities is as follows: Kuopio (KUO), Belfast (BFS), Turku (TKU), Grenoble (GNB), Luxembourg (LUX), Canberra (CBR), Palermo (PMO), Nantes (NTE), Rennes (RNS), Detroit (DTW), Venice (VCE), Toulouse (TLS), Bordeaux (BOD), Winnipeg (YWG), Dublin (DUB), Adelaide (ADL), Brisbane (BNE), Lisbon (LIS), Prague (PRG), Berlin (BER), Melbourne (MEL), Helsinki (HEL), Sydney (SYD), Rome (FCO), Paris (CDG).

The researchers collected General Transit Feed Specification (GTFS) from various public transport agencies, which provide their data in open-source. For walking connections, they used information from OpenStreetMap. They compute pair-to-pair connections, limiting walking connections to less than 1km distance. In our experiments, we implemented a more stringent walking limit of 600 meters due to memory errors encountered while computing the TCH-graph without these limitations. As an average walking speed, we took 1 meter per second. 

In our experiment, we used information about public transport schedules on extraction date provided in the dataset~\cite{dataset}.

\begin{table*}[t]
    \centering
    \caption{Build Time (Seconds), where TCH is Time-dependent Contraction Hierarchy, TTN-CST is Time Table Nodes with Combined Search Tree, TTN-FC is Time Table nodes with Fractional Cascading, ASC is ascending sorting of Fractional Cascading, DSC is descending sorting of Fractional Cascading, CHS is Contraction Hierarchies sorting of Fractional Cascading  }\label{tab:prec_time}
\renewcommand{\arraystretch}{1.2} 
 \tiny
    \setlength{\tabcolsep}{0.5pt} 
    \begin{tabular}
{|l|r@{~}|r@{~}|r@{~}|r@{~}|r@{~}|r@{~}|r@{~}|r@{~}|}
        \hline
        \multirow{4}{*}{\textbf{City}} & \textbf{TCH} & \multicolumn{7}{|c|}{\textbf{TTN}} \\  
        \cline{3-9}
        & & \multicolumn{3}{|c|}{\textbf{Original graph}} & \multicolumn{4}{c|}{\textbf{TCH}} \\
        \cline{3-9}
           && \textbf{TTN-CST} & \multicolumn{2}{|c|}{\textbf{TTN-FC}}  & \textbf{TTN-CST} & \multicolumn{3}{c|}{\textbf{TTN-FC}} \\
        \cline{4-5} \cline{7-9}
          &  && \textbf{ASC}  & \textbf{DSC}  & & \textbf{ASC}& \textbf{DSC} & \textbf{CHS}\\
        \hline
        KUO &  7 & 0 &  0 & 0 & 1 & 0 & 0 & 0 \\
        LUX &  57 & 0 & 0  & 0 & 7 & 4 & 4 & 5\\
        RNS &  5717 & 0 & 0  & 0 & 21 & 12 & 10 & 13 \\
        TKU&  591 & 0 &  0 & 0 & 42 & 22 &24  & 18\\
        VCE &  154 & 0 & 0  & 0 & 10 & 7 & 8 & 6\\
        BFS &  1611 & 0 &  0 & 0 & 30 & 18 & 20 & 14 \\
        PMO & 10055  & 1 &  0 & 0 & 109 & 54 & 44 & 57\\
        GNB &  942 & 0 & 0  & 0 & 62 & 35 & 27 & 36 \\
        CBR &  75 & 0 &  0 & 0 & 11 & 4 & 6 & 7 \\
        NTE &  3975 & 0 &  0 & 0 & 79 & 33 & 49 & 36 \\
        BER & 357  & 3 & 2 & 2 & 35 & 16 & 16 & 35\\
        DUB & 30959  & 1 & 0  & 1 & 259 & 104 & 115 & 111 \\
        TLS & 2936  & 0 &  0 & 0 & 123 & 45 & 67  & 48 \\
        YWG &  6548 & 1 & 0  & 1 & 351 & 135 & 154 & 146\\
        BOD & 18863  & 1 &  0 & 0 &  200 & 100 & 108 & 78 \\
        PRG & 3594  & 2 & 1  & 1 & 322 & 163 & 160 & 170 \\
        DTW & 6680  & 0 & 0  & 0 & 167 & 46 & 48 & 47\\
        LIS & 7841  & 2 & 0  & 1 & 478 & 104 & 521 & 532\\
        HEL & 6287  & 2 & 1  & 1 & 477 & 107 & 115 & 574 \\
         ADL & 2242  & 1 & 0  & 0 & 188 & 95 &  97 & 66 \\
         FCO & 53179  & 5 &  2 & 3 & 1238 & 1508 & 1186 & 1257 \\
         BNE & 5123  & 1 &  0 & 1 & 396 & 159 & 111  & 159\\
         CDG & --  &  8 &  4  & 5 & -- & -- & -- & -- \\
         MEL &  18009 & 4  &  1 & 1 & 748 & 451 & 486 & 320 \\
         SYD & 27128  & 5 & 1 & 1 & 1259 & 968 &1510  & 855 \\

        \hline
    \end{tabular}
\end{table*}

Table~\ref{tab:stats} contains various parameters related to the city's graphs.

\begin{table*}[t]
    \centering
    \caption{Comparing the graphs: Total size of merged timetable (\#$C_{m}$),  nodes (\#N), number edges (\# E), the mean amount of outgoing edges (Mean E), the mean amount of outgoing edges, which contain timetable information (Mean TT E), the mean size of timetables in edges (Mean TT) and percentage of ATF functions with timetables (\% TT). 
 Results shown for the original graph \textbf{---} and the contracted graph TCH.}\label{tab:stats}
\renewcommand{\arraystretch}{1.2} 
 \tiny
    \setlength{\tabcolsep}{0.5pt} 
    \begin{tabular}
{|l|r@{~}|r@{~}|r@{~}|r@{~}|r@{~}|r@{~}|r@{~}|r@{~}|r@{~}|r@{~}|r@{~}|r@{~}|}
        \hline
        \multirow{3}{*}{} & \multirow{3}{*}{}& \multirow{3}{*}{} & \multicolumn{2}{c|}{\textbf{\# E}}& \multicolumn{2}{c|}{\textbf{Mean E}} & \multicolumn{2}{c|}{\textbf{Mean TT E}} & \multicolumn{2}{c|}{\textbf{Mean TT}} &
        \multicolumn{2}{c|}{\textbf{\% TT}}\\
        
        \cline{4-13} 
        
        \textbf{City} & \textbf{\#$C_{m}$}& \textbf{\#N} & \textbf{---} & \textbf{TCH}& \textbf{---} & \textbf{TCH} & \textbf{---} & \textbf{TCH} & \textbf{---} & \textbf{TCH} & \textbf{---} & \textbf{TCH}\\
        \hline
        KUO &32122& 549 & 3994 & 8320 & 7.28 & 15.15 & 1.28 & 9.83 & 41.85 & 58.07 & 18\% & 65\%  \\
        LUX & 186752 & 1367 & 6389 & 17129 & 4.67 & 12.53 & 2.36 & 10.56 & 53.93 & 68.72 & 50\% & 84\%  \\
        RNS & 109075 & 1407 & 12128 & 31409 & 8.62 & 22.32 & 1.19 & 15.55 & 63.98 & 108.52 & 14\% & 70\%  \\
        TKU & 133512 & 1850 & 19971 & 51986 & 10.80 & 28.10 & 1.26 & 18.84 & 55.82 & 119.50 & 12\% & 67\%  \\
        VCE & 118519 & 1874 & 13365 & 33100 & 7.13 & 17.66 & 1.44 & 12.78 & 41.44 & 61.45 & 20\% & 72\%  \\
        
BFS & 122693 & 1917 & 25040 & 65372 & 13.06 & 34.10 & 1.14 & 22.99 & 55.16 & 84.64 & 9\% & 67\%  \\
PMO & 226215 & 2176 & 35268 & 103006 & 16.21 & 47.34 & 1.18 & 32.61 & 86.67 & 140.42 & 7\% & 69\%  \\
GNB & 114492 & 2231 & 29559 & 81308 & 13.25 & 36.44 & 0.75 & 23.84 & 66.96 & 124.73 & 6\% & 65\%  \\
CBR & 124305 & 2767 & 22127 & 56379 & 8.00 & 20.38 & 1.16 & 14.37 & 37.37 & 44.67 & 15\% & 71\%  \\

NTE & 196421 & 3430 & 46928 & 127683 & 13.68 & 37.23 & 0.81 & 24.66 & 69.55 & 104.20 & 6\% & 66\%  \\
BER & 1048218 & 4601 & 17908 & 59997 & 3.89 & 13.04 & 2.52 & 11.91 & 86.96  & 108.69 & 65\% & 91\% \\
DUB & 407240 & 4620 & 51868 & 214719 & 11.23 & 46.48 & 1.20 & 36.70 & 70.09 & 115.12 & 11\% & 79\%  \\
TLS & 224516 & 4969 & 69185 & 196561 & 13.92 & 39.56 & 0.76 & 26.51 & 58.34 & 91.91 & 5\% & 67\%  \\
YWG & 333882 & 5079 & 81794 & 335158 & 16.10 & 65.99 & 1.15 & 51.68 & 56.25 & 98.54 & 7\% & 78\%  \\
BOD & 236595 & 5288 & 85458 & 269689 & 16.16 & 51.00 & 0.76 & 35.23 & 58.17 & 104.65 & 5\% & 69\%  \\
PRG & 670423 & 5485 & 45452 & 199003 & 8.29 & 36.28 & 1.23 & 29.37 & 92.07  & 146.59 & 14\%& 81\% \\
DTW & 214863 & 5683 & 80110 & 381354 & 14.10 & 67.10 & 1.05 & 54.44 & 36.10 & 59.21 & 7\% & 81\%  \\
LIS & 526179 & 7073 & 93533 & 323498 & 13.22 & 45.74 & 1.27 & 34.81 & 57.55 & 128.85 & 10\% & 76\%  \\
HEL & 686457 & 7077 & 85394 & 333752 & 12.07 & 47.16 & 1.28 & 37.12 & 71.5 & 125.50 & 11\% & 79\%  \\
ADL & 404300 & 7662 & 72154 & 276195 & 9.42 & 36.05 & 1.21 & 28.43 & 43.24 & 69.05 & 13\% & 79\%  \\
FCO & 1051211 & 7869 & 110647 & 553727 & 14.06 & 70.37 & 1.29 & 57.89 & 103.5 & 162.33 & 9\% & 82\%  \\
BNE & 392805 & 9802 & 103302 & 461558 & 10.54 & 47.09 & 1.20 & 38.13 & 32.17 & 65.43 & 11\% & 81\%  \\

CDG & 1823872 & 12290 & 193723 & --- & 15.76 & --- & 1.19 & --- & 121.28 & --- & 8\% & ---  \\

        MEL & 1098227 & 19493 & 200621 & 887352 & 10.29 & 45.52 & 1.11 & 36.95 & 49.81 & 82.05 & 11\% & 81\%  \\
        SYD & 1265135 & 24261 & 286150 & 1464458 & 11.79 & 60.36 & 1.19 & 50.47 & 41.94 & 65.57 & 10\% & 84\%  \\

        \hline
    \end{tabular}
\end{table*}

Table~\ref{tab:mem} shows the memory usage of different graphs we used in our research and information about space costs, which is needed for different precomputation techniques. Unfortunately, we could not precompute CH-graph in Paris due to memory issues, so these results are omitted.

\begin{table*}[t]
    \centering
    \caption{Memory usage comparison for Basic graph and TCH-graph (in MB), where TTE is timetable edges (the standard approach, where we run evaluation on the edge level), TTN-CST is Time Table Nodes with Combined Search Tree, TTN-FC is Time Table nodes with Fractional Cascading, ASC is ascending sorting of Fractional Cascading, DSC is descending sorting of Fractional Cascading, CHS is Contraction Hierarchies sorting of Fractional Cascading}\label{tab:mem} 
    \renewcommand{\arraystretch}{1.2} 
    \tiny
    \setlength{\tabcolsep}{0.01pt} 
    \begin{tabular}
{|l|r@{~}|r@{~}|r@{~}|r@{~}|r@{~}|r@{~}|r@{~}|r@{~}|r@{~}|r@{~}|}
        \hline
        \multirow{4}{*}{\textbf{City}}  &\multirow{4}{*}{\textbf{CSA}} &\multicolumn{4}{|c|}{\textbf{Original graph}} & \multicolumn{5}{c|}{\textbf{TCH}} \\
        \cline{3-11}
          & & \textbf{TTE} & \textbf{TTN-CST} & \multicolumn{2}{|c|}{\textbf{TTN-FC}}  & \textbf{TTE} & \textbf{TTN-CST} & \multicolumn{3}{c|}{\textbf{TTN-FC}} \\
        \cline{5-6} \cline{9-11}
          &  & & & \textbf{ASC}  & \textbf{DSC} & & & \textbf{ASC}& \textbf{DSC} & \textbf{CHS}\\
        \hline
        KUO & 1 & 1 & 2 & 1 & 1 & 22 & 33 & 27 & 27 & 28  \\ 
LUX & 7 & 7 & 13 & 10 & 10 & 78 & 125 & 96 & 97 & 98  \\ 
RNS & 4 & 4 & 11 & 6 & 6 & 181 & 299 & 222 & 224 & 223  \\ 
TKU & 5 & 7 & 17 & 9 & 9 & 320 & 564 & 391 & 393 & 391  \\ 
VCE & 4 & 5 & 11 & 7 & 7 & 118 & 188 & 143 & 145 & 145  \\ 
BFS & 4 & 6 & 15 & 8 & 8 & 341 & 530 & 401 & 403 & 403  \\ 
PMO & 8 & 10 & 31 & 14 & 14 & 882 & 1470 & 1053 & 1057 & 1049  \\ 
GNB & 4 & 6 & 16 & 7 & 7 & 555 & 894 & 661 & 664 & 660  \\ 
CBR & 4 & 6 & 12 & 8 & 8 & 164 & 245 & 194 & 196 & 197  \\ 
NTE & 7 & 10 & 27 & 13 & 13 & 799 & 1242 & 943 & 947 & 942  \\ 
BER & 41 & 41 & 71 & 58 & 60 & 621 & 809 & 723 & 728 & 730  \\ 
DUB & 15 & 18 & 47 & 24 & 24 & 2066 & 3408 & 2402 & 2408 & 2396  \\ 
TLS & 8 & 13 & 32 & 16 & 16 & 1164 & 1850 & 1362 & 1367 & 1362  \\ 
YWG & 13 & 18 & 51 & 22 & 23 & 3044 & 4849 & 3485 & 3493 & 3466  \\ 
BOD & 9 & 14 & 38 & 18 & 18 & 1873 & 2979 & 2193 & 2198 & 2182  \\ 
PRG & 25 & 27 & 66 & 36 & 36 & 2505 & 3924 & 2910 & 2920 & 2921  \\ 
DTW & 8 & 13 & 31 & 16 & 16 & 3046 & 4046 & 3358 & 3362 & 3344  \\ 
LIS & 20 & 26 & 74 & 34 & 34 & 3337 & 5600 & 3902 & 3913 & 3883  \\ 
HEL & 27 & 31 & 82 & 40 & 41 & 3706 & 5814 & 4282 & 4295 & 4262  \\ 
ADL & 16 & 20 & 46 & 26 & 26 & 1919 & 2940 & 2181 & 2188 & 2181  \\ 
FCO & 41 & 49 & 143 & 63 & 65 & 8722 & 14087 & 10024 & 10044 & 9946  \\ 
BNE & 15 & 21 & 50 & 27 & 28 & 2744 & 4904 & 3167 & 3175 & 3172  \\ 
CDG & 72 & 87 & 262 & 114 & 115 & --- & --- & --- & --- & --- \\
MEL & 43 & 56 & 132 & 72 & 72 & 10014 & 13654 & 11021 & 11036 & 10977  \\ 
SYD & 51 & 67 & 157 & 85 & 86 & 13139 & 19078 & 14512 & 14535 & 14503  \\ 

        \hline
    \end{tabular}
\end{table*}

It is essential to admit that in our realization we store string information about public transport in our graph, which we will use to move from one station to another. In the production-ready service, this data should be converted to an integer, and the name should be stored separately, reducing memory usage.

\subsubsection{Query Processing Time}

For comparing query times, we picked 1000 random combinations of stops and times between extractions dates in the dataset \cite{dataset}. 

Results in average runtime in microseconds for the different precomputation techniques are presented in Table~\ref{tab:run}.

In our dataset, we have connections with zero length, where the departure and arrival times are equal. The classical CSA algorithm cannot handle such cases. Therefore, we modified the CSA algorithm to reiterate each time all connections with zero length, in case some of the arrival times have been updated.

It is important to note that graph-based approaches can move multiple times by walking along edges. With CSA, we can only allow walking connections from one stop to neighbouring stops once, and we cannot repeat this operation from another stop that we reach by walking. This is why graph-based approaches sometimes find paths that CSA does not. In table \ref{tab:run}, we also include information about the percentage of paths that exist for each of the algorithms.

\begin{table*}[t]
    \centering
    \caption{Average runtime in ms, percentage of finding paths (\%P), average amount of expanded nodes (\#Exp), Performance of TCH-TTN-CST over the CSA (\%Perf), where TTE is timetable edges (the standard approach, where we run evaluation on the edge level), TTN-CST is Time Table Nodes with Combined Search Tree, TTN-FC is Time Table nodes with Fractional Cascading}\label{tab:run}
    \renewcommand{\arraystretch}{1.2} 
    \tiny
    \setlength{\tabcolsep}{0.01pt} 
    \begin{tabular}{|l|r@{~}|r@{~}|r@{~}|r@{~}|r@{~}|r@{~}|r@{~}|r@{~}|r@{~}|r@{~}|r@{~}|r@{~}|r@{~}|r@{~}|r@{~}|}
        \hline
        & \multicolumn{2}{c|}{} & \multicolumn{12}{c|}{\textbf{Graph-based}} & \\
        \cline{2-15}
         &  & & \multicolumn{5}{c|}{\textbf{Original graph}} & \multicolumn{6}{c|}{\textbf{TCH}} & & \\
        \cline{4-14}
         \textbf{City} & & \textbf{\%P} &\textbf{\#Exp} & \multicolumn{4}{c|}{\textbf{Avg runtime}}& \textbf{\#Exp} & \multicolumn{5}{c|}{\textbf{Avg runtime}} &  \textbf{\%P} & \textbf{\%Perf} \\
          \cline{5-8} \cline{10-14}
           &  \textbf{CSA} && & \textbf{TTE} & \textbf{TTN-CST} & \multicolumn{2}{c|}{\textbf{TTN-FC}} & & \textbf{TTE} & \textbf{TTN-CST} & \multicolumn{3}{c|}{\textbf{TTN-FC}} & & \\
        \cline{7-8} \cline{12-14}
          & && & & &\textbf{ASC} & \textbf{DSC} & & & & \textbf{ASC} & \textbf{DSC} & \textbf{CHS} & & \\
        \hline
        KUO & 1.0 & 79 & 282 & 2.7 & 2.4 & 2.8 & 2.3 & 52 & 1.7 & \textbf{0.9} & 1.4 & 1.4 & 1.2 & 84 & \textbf{111} \\
LUX & 5.2 & 87 & 734 & 10.9 & 7.3 & 10.7 & 7.2 & 75 & 3.7 & \textbf{2.2} & 3.5 & 3.6 & 3.0 & 89 & \textbf{236} \\
RNS & \textbf{2.8} & 82 & 668 & 12.1 & 8.8 & 8.1 & 7.7 & 116 & 11.4 & 5.9 & 8.7 & 8.6 & 7.5 & 84 & 47 \\
TKU & \textbf{4.4} & 82 & 995 & 17.3 & 16.5 & 15.1 & 14.6 & 146 & 19.1 & 12.3 & 18.1 & 17.6 & 14.6 & 90 & 36 \\
VCE & 5.0 & 69 & 792 & 10.8 & 9.7 & 9.3 & 9.3 & 85 & 5.1 & \textbf{3.2} & 4.5 & 4.4 & 4.2 & 74 &  \textbf{156}\\
BFS & \textbf{4.5} & 89 & 1116 & 27.0 & 23.6 & 23.9 & 23.6 & 154 & 18.3 & 10.3 & 16.6 & 15.3 & 13.3 & 99 &  44 \\
PMO & \textbf{5.7} & 89 & 1261 & 31.9 & 31.5 & 28.6 & 27.2 & 210 & 34.4 & 18.1 & 30.1 & 29.7 & 27.1 & 100 &  31\\
GNB & \textbf{4.1} & 72 & 1069 & 20.7 & 18.2 & 16.1 & 15.7 & 175 & 31.5 & 16.0 & 25.0 & 27.1 & 24.9 & 82 &  27\\
CBR & 6.4 & 82 & 1362 & 22.9 & 21.1 & 20.6 & 20.6 & 120 & 9.5 & \textbf{5.6} & 8.9 & 9.3 & 8.7 & 86 & \textbf{114} \\
NTE & \textbf{7.6} & 82 & 1792 & 42.0 & 40.9 & 38.5 & 38.6 & 199 & 32.6 & 19.1 & 28.0 & 29.0 & 26.1 & 92 &  40 \\
BER & 30.7 & 75 & 2291 & 34.8 & 28.3 & 34.8 & 33.4 & 123 & 16.8 & \textbf{8.3} & 15.4 & 14.9 & 13.1 & 82 &  \textbf{370}\\
DUB & \textbf{16.2} & 86 & 2529 & 53.1 & 55.1 & 53.6 & 48.3 & 329 & 108.8 & 59.2 & 86.1 & 87.5 & 77.8 & 96 &  27\\
TLS & \textbf{11.3} & 75 & 2549 & 60.5 & 55.2 & 51.3 & 49.3 & 262 & 61.1 & 34.7 & 51.0 & 51.3 & 46.0 & 88 & 33 \\
YWG & \textbf{14.7} & 85 & 2903 & 80.7 & 70.5 & 71.6 & 64.1 & 407 & 134.3 & 80.5 & 112.1 & 120.4 & 105.4 & 100 & 18 \\
BOD & \textbf{11.0} & 81 & 2690 & 70.4 & 62.1 & 53.7 & 54.9 & 311 & 83.7 & 46.8 & 70.4 & 72.9 & 64.1 & 93 &  24\\
PRG & \textbf{20.8} & 79 & 2477 & 44.2 & 44.7 & 45.1 & 43.7 & 284 & 83.7 & 46.5 & 68.4 & 69.7 & 63.4 & 81 &  45\\
DTW & \textbf{11.5} & 80 & 3110 & 75.1 & 73.7 & 72.1 & 69.3 & 403 & 111.0 & 68.8 & 98.7 & 100.3 & 92.7 & 100 & 17 \\
LIS & \textbf{27.2} & 76 & 3353 & 85.5 & 91.9 & 83.2 & 81.6 & 335 & 126.7 & 74.8 & 105.8 & 106.8 & 89.4 & 83 & 36 \\
HEL & \textbf{26.4} & 82 & 3708 & 83.2 & 74.1 & 73.1 & 69.5 & 381 & 121.3 & 74.2 & 104.4 & 104.0 & 93.6 & 89 & 36 \\
ADL & \textbf{22.5} & 79 & 4061 & 77.5 & 69.5 & 64.8 & 64.9 & 327 & 84.4 & 47.4 & 71.3 & 73.5 & 67.6 & 92 &  47\\
FCO & \textbf{30.2} & 83 & 4185 & 102.3 & 102.9 & 92.4 & 89.6 & 540 & 306.9 & 522.3 & 807.1 & 897.8 & 824.1 & 93 & 6 \\
BNE & \textbf{28.6} & 71 & 4855 & 92.8 & 83.2 & 79.1 & 79.6 & 467 & 196.8 & 115.9 & 163.7 & 165.6 & 142.8 & 86 & 25 \\
CDG & \textbf{51.7}& 87 & 7154 & 186.2 & 178.8 & 163.7 & 169.2 &-- & --&--&--&--&--& 98 & -- \\
MEL & \textbf{70.1} & 71 & 9585 & 211.1 & 231.5 & 214.2 & 215.8 & 587 & 267.5 & 478.1 & 680.5 & 713.5 & 694.3 & 88 & 15 \\
SYD & \textbf{130.0} & 67 & 12404 & 313.8 & 328.4 & 302.9 & 311.6 & 788 & 604.2 & 1886.1 & 3883.7 & 4349.0 & 3632.7 & 84 & 7 \\
        \hline
    \end{tabular}
\end{table*}

As we can see from Table~\ref{tab:run}, CSA performs better for most cities and takes less memory. In 20\% of cities, the opposite occurred: Kuopio, Luxembourg, Venice, Canberra, and Berlin. These cities are characterized by a small number of outgoing edges in the original graph (see Figure ~\ref{fig:mean_e}) and a high percentage of Timetable edges in the original graph (see Figure ~\ref{fig:perc_tt}).

\pgfplotsset{colormap={custom}{
    color=(blue) color=(cyan) color=(green) color=(lime) color=(yellow) color=(orange) color=(red)
}}

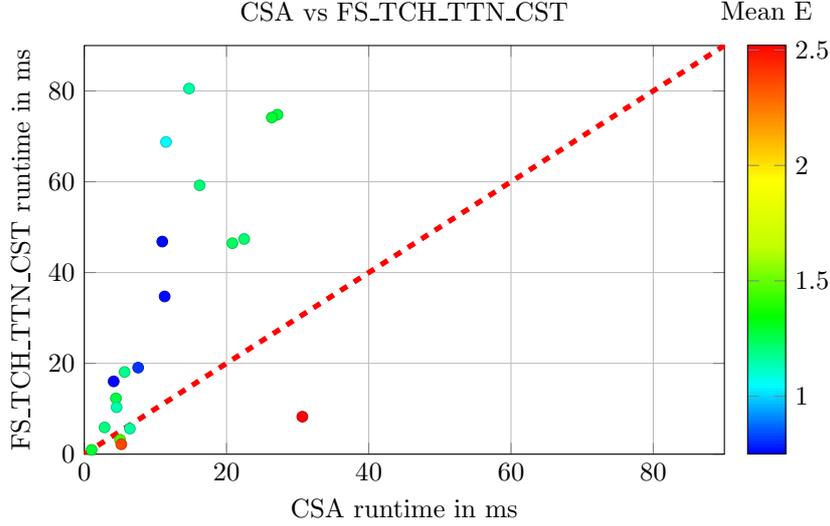
\begin{figure}[h!]
    \centering
    \begin{tikzpicture}
        \begin{axis}[
            title={CSA vs FS\_TCH\_TTN\_CST},
            xlabel={CSA runtime in ms},
            ylabel={FS\_TCH\_TTN\_CST runtime in ms},
            xmin=0, xmax=90,
            ymin=0, ymax=90,
            legend style={at={(1,0.75)}, anchor=south east},
            grid=major,
            width=10cm,
            height=7cm,
            x axis line style={-},
            y axis line style={-},
            colormap name=custom, 
            colorbar, 
            colorbar style={title={Mean E}},
        ]
            \addplot[
                scatter,
                only marks,
                scatter src=explicit,
                mark=*
            ]
            coordinates {
                (1.029, 0.944) [1.28]
                (2.846, 5.886) [1.19]
                (4.448, 12.298) [1.26]
                (5.017, 3.174) [1.44]
                (4.538, 10.33) [1.14]
                (5.661, 18.099) [1.18]
                (4.125, 16.031) [0.75]
                (6.403, 5.626) [1.16]
                (7.571, 19.062) [0.81]
                (30.656, 8.257) [2.52]
                (16.224, 59.208) [1.20]
                (11.298, 34.738) [0.76]
                (14.737, 80.52) [1.15]
                (10.966, 46.823) [0.76]
                (20.825, 46.459) [1.23]
                (11.493, 68.763) [1.05]
                (27.157, 74.768) [1.27]
                (26.366, 74.153) [1.28]
                (22.479, 47.39) [1.21]
                (30.21, 522.26) [1.29]
                (28.643, 115.905) [1.20]
                (70.067, 478.115) [1.11]
                (130.006, 1886.081) [1.19]
                (5.201, 2.179) [2.36]
            };

            \addplot[
                color=red,
                line width=2pt,
                dashed
            ]
            coordinates {(0, 0) (150, 150)};
        \end{axis}
    \end{tikzpicture}
    \caption{FS\_TCH\_TTN\_CST vs CSA runtime colored by Mean $E$. Visualization provides just samples, where the runtime of FS\_TCH\_TTN\_CST is smaller than 90 ms.}
    \label{fig:mean_e}
\end{figure}

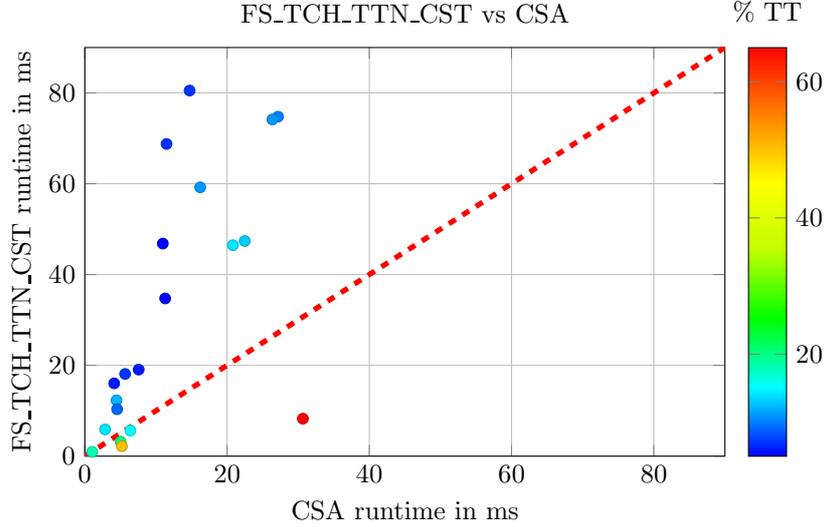
\begin{figure}[h!]
    \centering
    \begin{tikzpicture}
        \begin{axis}[
            title={FS\_TCH\_TTN\_CST vs CSA},
            xlabel={CSA runtime in ms},
            ylabel={FS\_TCH\_TTN\_CST runtime in ms},
            xmin=0, xmax=90,
            ymin=0, ymax=90,
            legend style={at={(1,0.75)}, anchor=south east},
            grid=major,
            width=10cm,
            height=7cm,
            x axis line style={-},
            y axis line style={-},
            colormap name=custom, 
            colorbar, 
            colorbar style={title={\% TT}},
        ]
            \addplot[
                scatter,
                only marks,
                scatter src=explicit,
                mark=*
            ]
            coordinates {
                (1.029, 0.944) [18]
                
                (2.846, 5.886) [14]
                (4.448, 12.298) [12]
                (5.017, 3.174) [20]
                (4.538, 10.33) [9]
                (5.661, 18.099) [7]
                (4.125, 16.031) [6]
                (6.403, 5.626) [15]
                (7.571, 19.062) [6]
                (30.656, 8.257) [65]
                (16.224, 59.208) [11]
                (11.298, 34.738) [5]
                (14.737, 80.52) [7]
                (10.966, 46.823) [5]
                (20.825, 46.459) [14]
                (11.493, 68.763) [7]
                (27.157, 74.768) [10]
                (26.366, 74.153) [11]
                (22.479, 47.39) [13]
                (30.21, 522.26) [9]
                (28.643, 115.905) [11]
                (70.067, 478.115) [11]
                (130.006, 1886.081) [10]
                (5.201, 2.179) [50]
            };

            \addplot[
                color=red,
                line width=2pt,
                dashed
            ]
            coordinates {(0, 0) (150, 150)};
        \end{axis}
    \end{tikzpicture}
    \caption{FS\_TCH\_TTN\_CST vs CSA runtime colored by percentage of ATF functions with timetables (\% TT) . Visualization provides just samples, where the runtime of FS\_TCH\_TTN\_CST is smaller than 90 ms.}
    \label{fig:perc_tt}
\end{figure}

We also want to admit that TTN improves the algorithm's performance for all graph-based approaches. An interesting observation is that after some level of complexity, CH becomes an unnecessary precomputation in the case of Multimodal Pathfinding with Arrival Time Functions. The same things have been mentioned by Marty and Viennot ~\cite{contraction_hierarchy_transport_oliver}. 
We want to admit that the reason for this is an increase in the binary search calculation process due to the growth of the count of outgoing edges and the complexity of the newly inserted shortcuts. In Table~\ref{tab:run}, we can observe that Dijkstra's average number of expanded nodes during the pathfinding process is still much larger than that of Forward Search.

We hypothesise that preprocessing techniques do not work for high-density graphs, and more straightforward algorithms like CSA work better because there are fewer binary search operations during the search process. TTN is a step forward in simplifying this process in graph-based algorithms. We observed that in cases where we have a low number of outgoing edges, a graph-based structure performs still better than CSA, and TTN techniques help to reduce the complexity of the node evaluation process. An interesting situation has arisen in a city like Venice or Canberra, where using TCH conventionally does not offer any benefits over the CSA search, but applying TTN shows advantages.

An interesting case also appears in Berlin. For this city, the improvement in TCH-TTN over the CSA is the highest, equal to 3.7. We assume that this is related to the fact that this city has the highest percentage of outgoing timetable edges (see Figure ~\ref{fig:perc_tt}). We also note that the degree of improvement is correlated with the proportion of the edges with timetable information.

The ASC and DSC ordering does not seem to have a significant difference, but the CHS method shows better results, especially with smaller graphs. Further research is needed to explore the impact of list order in FC.

For the graph-based approaches, TTN-CST improves Forward Search speed performance by 1.8 times (if not considering large graphs). And TTN-FC speeds up Dijkstra by 1.1 times. The level of improvement depends on the graph density. We noticed that in the original graph, TTN-FC performed better than TTN-CST. However, in the TCH situation, the opposite is true. We decided to explore and analyze that thing in more detail in the next session by building a synthetic graph for it.

TTN-CST precomputation increases the memory space by 2.4 times in the Basic Transport Graph and 1.6 times in the TCH graph. The space complexity of such a technique is $\mathcal{O} (k^2 \times |C| \times n)$, where $k$ is the average count of outgoing edges, $|C|$ is the average complexity of the timetable, and $n$ is the number of nodes in a graph. Meanwhile, the space complexity of TTN-FC is smaller and equal to $\mathcal{O}(k\times|C| \times n)$. In practice, TTN-FC increases memory by just 1.29 times for the Basic Transport Graph and 1.17 times for the TCH graph.

\ignore{
\begin{table}[H]
\centering
\caption{Average number $\pm$ standard deviation of expanded nodes during the search}
\begin{tabular}{|p{1.9cm}|r@{}l|r@{}l|r@{}l|r@{}l|r@{}l|}
\hline
  &\multicolumn{2}{c|}{\textbf{KUO}} & \multicolumn{2}{c|}{\textbf{BER}} 
 & \multicolumn{2}{c|}{\textbf{PRG}} & \multicolumn{2}{c|}{\textbf{ADL}} & \multicolumn{2}{c|}{\textbf{BNE}} 
\\
\cline{2-11}
& \textbf{Avg} & \textbf{~Std} & \textbf{Avg} & \textbf{~Std} & \textbf{Avg} & \textbf{~Std} & \textbf{Avg} & \textbf{~Std} & \textbf{Avg} & \textbf{~Std}\\
\hline

  Dij. & 
 299  & $\pm$149 & 2365 & $\pm$279 & 2771 & $\pm$1505 & 3980 & $\pm$2134 & 5375 & $\pm$2636  \\ 
\hline

 FS over TCH &
 \textbf{58} & $\pm$\textbf{15} &  \textbf{137} & $\pm$\textbf{22} &  \textbf{320} & $\pm$\textbf{ 67} &  \textbf{340} & $\pm$\textbf{66} & \textbf{516} & $\pm$\textbf{106} \\  
\hline

\end{tabular}
\end{table}
}
\subsection{Synthetic graph analysis}

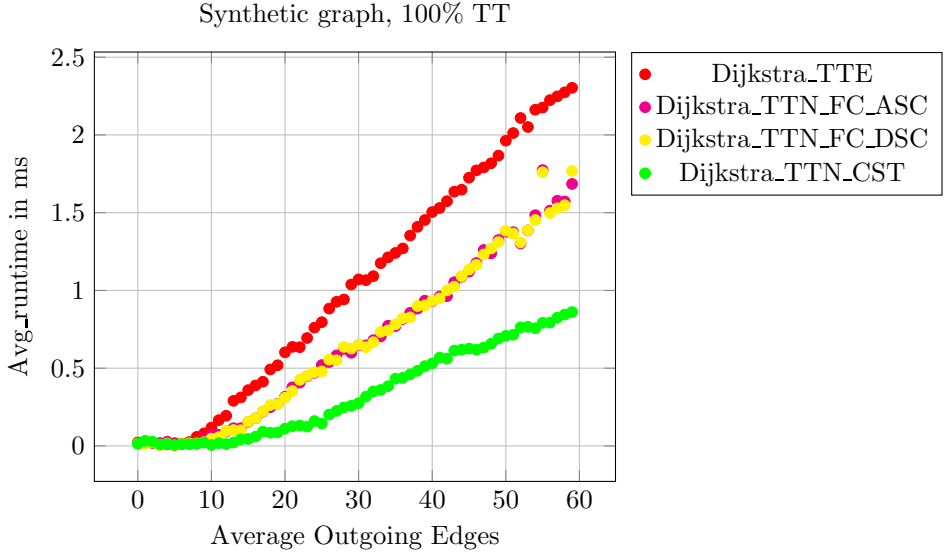
\begin{figure}
\centering
\begin{tikzpicture}
\begin{axis}[
    title={Synthetic graph, 100\% TT},
    xlabel={Average Outgoing Edges},
    ylabel={Avg\_runtime in ms},
    legend pos=outer north east,
    grid=both,
]

\addplot[
    only marks,
    mark=*,
    color=red,
    ]
    coordinates {
        (39, 1.453) (18, 0.492) (3, 0.018) (54, 2.162) (38, 1.409)
        (32, 1.092) (17, 0.413) (10, 0.118) (47, 1.791) (19, 0.519)
        (50, 1.963) (13, 0.29) (1, 0.03) (30, 1.071) (58, 2.274)
        (6, 0.015) (45, 1.725) (16, 0.389) (11, 0.166) (59, 2.303)
        (43, 1.635) (51, 2.012) (26, 0.883) (37, 1.353) (57, 2.248)
        (5, 0.018) (48, 1.818) (21, 0.637) (41, 1.53) (28, 0.942)
        (20, 0.602) (7, 0.025) (22, 0.635) (56, 2.224) (12, 0.194)
        (25, 0.796) (0, 0.022) (55, 2.176) (8, 0.058) (29, 1.038)
        (9, 0.081) (31, 1.066) (53, 2.051) (14, 0.312) (49, 1.867)
        (27, 0.927) (23, 0.694) (4, 0.01) (44, 1.648) (35, 1.241)
        (40, 1.504) (15, 0.359) (46, 1.772) (42, 1.574) (24, 0.761)
        (34, 1.213) (36, 1.27) (52, 2.109) (33, 1.175) (2, 0.019)
    };
\addlegendentry{Dijkstra\_TTE}

\addplot[
    only marks,
    mark=*,
    color=magenta,
    ]
    coordinates {
        (7, 0.02) (14, 0.114) (1, 0.022) (5, 0.013) (10, 0.071)
        (51, 1.375) (47, 1.26) (31, 0.647) (52, 1.302) (28, 0.624)
        (40, 0.93) (22, 0.407) (24, 0.47) (3, 0.016) (17, 0.22)
        (8, 0.015) (0, 0.019) (11, 0.072) (46, 1.176) (54, 1.484)
        (21, 0.377) (32, 0.68) (13, 0.114) (19, 0.271) (33, 0.704)
        (2, 0.021) (50, 1.377) (55, 1.774) (23, 0.448) (29, 0.599)
        (37, 0.856) (42, 0.963) (30, 0.647) (4, 0.027) (56, 1.513)
        (38, 0.883) (20, 0.316) (36, 0.816) (6, 0.014) (45, 1.121)
        (27, 0.582) (25, 0.52) (44, 1.088) (59, 1.685) (43, 1.053)
        (49, 1.324) (48, 1.238) (16, 0.178) (58, 1.572) (18, 0.249)
        (53, 1.386) (35, 0.771) (9, 0.041) (34, 0.773) (39, 0.934)
        (41, 0.96) (57, 1.577) (12, 0.093) (26, 0.54) (15, 0.155)
    };
\addlegendentry{Dijkstra\_TTN\_FC\_ASC}

\addplot[
    only marks,
    mark=*,
    color=yellow,
    ]
    coordinates {
        (18, 0.263) (58, 1.548) (1, 0.013) (24, 0.474) (27, 0.551)
        (29, 0.626) (37, 0.826) (39, 0.904) (15, 0.157) (42, 1.0)
        (32, 0.667) (9, 0.024) (21, 0.353) (53, 1.387) (51, 1.366)
        (10, 0.047) (0, 0.015) (7, 0.016) (28, 0.637) (38, 0.9)
        (45, 1.134) (33, 0.733) (47, 1.231) (14, 0.101) (12, 0.096)
        (50, 1.384) (44, 1.092) (49, 1.311) (16, 0.179) (40, 0.933)
        (8, 0.017) (36, 0.821) (5, 0.002) (52, 1.309) (4, 0.015)
        (34, 0.743) (2, 0.022) (3, 0.005) (48, 1.27) (31, 0.635)
        (13, 0.102) (17, 0.222) (26, 0.555) (46, 1.165) (41, 0.949)
        (55, 1.76) (19, 0.269) (11, 0.056) (23, 0.451) (25, 0.479)
        (20, 0.31) (43, 1.024) (35, 0.782) (57, 1.528) (59, 1.768)
        (6, 0.014) (30, 0.651) (56, 1.498) (54, 1.453) (22, 0.425)
    };
\addlegendentry{Dijkstra\_TTN\_FC\_DSC}

\addplot[
    only marks,
    mark=*,
    color=green,
    ]
    coordinates {
        (44, 0.619) (53, 0.766) (30, 0.275) (33, 0.36) (52, 0.762)
        (48, 0.657) (5, 0.006) (18, 0.085) (15, 0.045) (2, 0.027)
        (50, 0.708) (8, 0.014) (39, 0.513) (40, 0.531) (7, 0.011)
        (43, 0.613) (22, 0.131) (42, 0.561) (41, 0.568) (19, 0.087)
        (54, 0.757) (59, 0.861) (14, 0.043) (0, 0.014) (9, 0.021)
        (28, 0.248) (51, 0.715) (56, 0.793) (32, 0.351) (26, 0.202)
        (31, 0.317) (23, 0.124) (12, 0.012) (49, 0.691) (34, 0.384)
        (37, 0.46) (27, 0.225) (21, 0.127) (55, 0.792) (10, 0.006)
        (58, 0.845) (1, 0.032) (24, 0.16) (16, 0.06) (36, 0.438)
        (29, 0.258) (35, 0.432) (46, 0.619) (13, 0.022) (4, 0.012)
        (25, 0.144) (47, 0.633) (17, 0.092) (20, 0.111) (38, 0.483)
        (6, 0.011) (3, 0.01) (57, 0.825) (45, 0.626) (11, 0.018)
    };
\addlegendentry{Dijkstra\_TTN\_CST}

\end{axis}
\end{tikzpicture}
\caption{Synthetic graph with  100\% of edges with timetables (TT)}
\label{fig:alg_performance}
\end{figure}

To further investigate why TTN-CST performs better for TCH and TTN-FC performs better for the original graph, we created a synthetic random graph with 60 nodes and varying numbers of outgoing edges. In the first scenario, we established each edge to have both a walking connection and a transport connection. Specifically, we chose the connection in Sydney City between "High St Near Prince Of Wales Hospital" and "Coogee Bay Rd Near Perouse Rd". The timetable representing this connection has a size of 59, with a walking duration of 404 seconds. We build our experiment applying different modification of Dijkstra algorithm. As we varied the number of edges in the graph, we observed a correlation between duration time and algorithm selection, as depicted in Figure \ref{fig:alg_performance}. 

In the chart Figure \ref{fig:alg_performance}, we can see that TTN-CST outperforms other solutions, and the more outgoing edges we have, the better the performance. This situation is similar to what we observe for the TCH graph in Table 1, except for large graphs larger than 10Gb, for which each preprocessing technique does not yield significant results. We note that TTN-FC-ASC and TTN-FC-DSC produce the same results, which is expected since our graphs are synthetic with the same timetable.

\begin{figure}
\centering
\begin{tikzpicture}
\begin{axis}[
    title={Synthetic graph, 20\% TT},
    xlabel={Average Outgoing Edges},
    ylabel={Avg\_runtime in ms},
    legend pos=outer north east,
    grid=both,
]

\addplot[
    only marks,
    mark=*,
    color=red,
    ]
    coordinates {
        (26, 0.662) (17, 0.341) (52, 1.597) (1, 0.014) (10, 0.017) (13, 0.157) (19, 0.41) (38, 1.097) (39, 1.16) (8, 0.014)
        (9, 0.01) (15, 0.245) (56, 1.788) (29, 0.747) (35, 0.994) (27, 0.686) (14, 0.192) (6, 0.007) (25, 0.629) (55, 1.803)
        (31, 0.833) (54, 1.735) (7, 0.014) (30, 0.79) (34, 0.968) (53, 1.688) (2, 0.013) (37, 1.091) (40, 1.187) (36, 1.013)
        (21, 0.471) (49, 1.531) (23, 0.528) (33, 0.893) (18, 0.36) (5, 0.002) (57, 1.77) (16, 0.282) (24, 0.555) (43, 1.264)
        (3, 0.012) (4, 0.008) (11, 0.061) (32, 0.855) (42, 1.28) (0, 0.015) (12, 0.108) (48, 1.483) (41, 1.211) (46, 1.402)
        (28, 0.744) (50, 1.567) (58, 1.79) (22, 0.511) (44, 1.277) (51, 1.589) (20, 0.437) (59, 1.844) (45, 1.35) (47, 1.46)
    };
\addlegendentry{Dijkstra\_TTE}

\addplot[
    only marks,
    mark=*,
    color=magenta,
    ]
    coordinates {
        (30, 0.098) (27, 0.054) (15, 0.006) (57, 0.51) (19, 0.006) (29, 0.081) (52, 0.432) (33, 0.131) (56, 0.505) (2, 0.012)
        (9, 0.001) (11, 0.002) (36, 0.22) (24, 0.011) (6, 0.0) (3, 0.002) (14, 0.009) (46, 0.362) (53, 0.477) (5, 0.002)
        (22, 0.017) (59, 0.546) (54, 0.515) (8, 0.007) (35, 0.184) (45, 0.352) (32, 0.12) (31, 0.113) (10, 0.005) (58, 0.553)
        (23, 0.014) (39, 0.259) (0, 0.012) (18, 0.009) (28, 0.052) (21, 0.013) (47, 0.369) (20, 0.011) (16, 0.005) (41, 0.305)
        (17, 0.004) (25, 0.026) (13, 0.008) (42, 0.321) (38, 0.237) (1, 0.006) (50, 0.422) (55, 0.528) (43, 0.339) (49, 0.382)
        (44, 0.359) (34, 0.178) (26, 0.034) (7, 0.002) (12, 0.006) (37, 0.214) (48, 0.393) (51, 0.43) (4, 0.002) (40, 0.281)
    };
\addlegendentry{Dijkstra\_TTN\_FC\_ASC}

\addplot[
    only marks,
    mark=*,
    color=yellow,
    ]
    coordinates {
        (17, 0.011) (20, 0.016) (38, 0.234) (39, 0.257) (31, 0.107) (23, 0.013) (21, 0.007) (59, 0.539) (36, 0.206) (32, 0.112)
        (0, 0.004) (33, 0.136) (56, 0.477) (41, 0.307) (43, 0.323) (10, 0.005) (9, 0.005) (40, 0.287) (45, 0.347) (46, 0.341)
        (28, 0.054) (54, 0.504) (35, 0.186) (25, 0.03) (47, 0.369) (37, 0.229) (3, 0.001) (8, 0.009) (1, 0.009) (18, 0.008)
        (6, 0.002) (34, 0.16) (14, 0.006) (24, 0.024) (58, 0.554) (42, 0.307) (52, 0.436) (44, 0.345) (48, 0.404) (16, 0.002)
        (22, 0.009) (30, 0.093) (19, 0.008) (49, 0.374) (57, 0.496) (29, 0.062) (5, 0.002) (26, 0.024) (13, 0.003) (4, 0.004)
        (51, 0.411) (27, 0.044) (12, 0.001) (7, 0.002) (15, 0.006) (53, 0.484) (11, 0.003) (50, 0.421) (55, 0.512) (2, 0.005)
    };
\addlegendentry{Dijkstra\_TTN\_FC\_DCS}

\addplot[
    only marks,
    mark=*,
    color=green,
    ]
    coordinates {
        (25, 0.446) (27, 0.484) (37, 0.757) (14, 0.055) (24, 0.399) (53, 1.227) (4, 0.006) (26, 0.445) (6, 0.007) (32, 0.587)
        (10, 0.004) (54, 1.284) (48, 1.097) (47, 1.012) (40, 0.866) (23, 0.356) (19, 0.237) (39, 0.795) (9, 0.014) (33, 0.617)
        (28, 0.513) (30, 0.532) (35, 0.705) (57, 1.299) (21, 0.273) (15, 0.097) (18, 0.197) (56, 1.303) (50, 1.123) (7, 0.004)
        (58, 1.302) (45, 0.978) (22, 0.323) (5, 0.006) (51, 1.164) (11, 0.012) (29, 0.539) (3, 0.006) (17, 0.179) (38, 0.78)
        (49, 1.109) (55, 1.292) (16, 0.117) (1, 0.006) (13, 0.034) (0, 0.017) (59, 1.382) (42, 0.894) (44, 0.937) (31, 0.581)
        (2, 0.008) (52, 1.159) (41, 0.88) (34, 0.674) (8, 0.001) (20, 0.255) (36, 0.73) (12, 0.024) (46, 1.003) (43, 0.903)
    };
\addlegendentry{Dijkstra\_TTN\_CST}

\end{axis}
\end{tikzpicture}
\caption{Synthetic graph, 20\% of ATF with timetables (TT)}
\label{fig:alg_performance_20}
\end{figure}
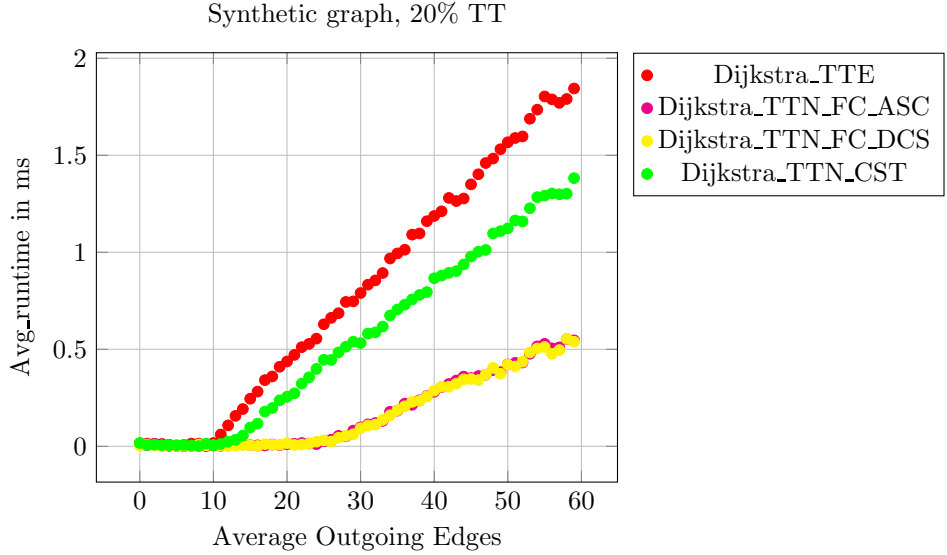

After this observation, the logical question is: under what circumstances does TTN-FC work faster than TTN-CST? To answer this question, we left all walking connections between the nodes but decreased the percentage of nodes with public transport connections to 20 \% and repeated the experiment. The results are depicted in the Figure \ref{fig:alg_performance_20}. 

In Figure \ref{fig:alg_performance_20}, we notice that as the average amount of outgoing edges increases, the TTN approach shows higher improvement. However, in this instance, TTN-CST demonstrates better performance. This trend is also evident in experiments involving real cities. In the original graph, TTN-FC outperformed the TCH graph, except in the case of Berlin, where the percentage of timetable outgoing edges in the original graph was 65\%.

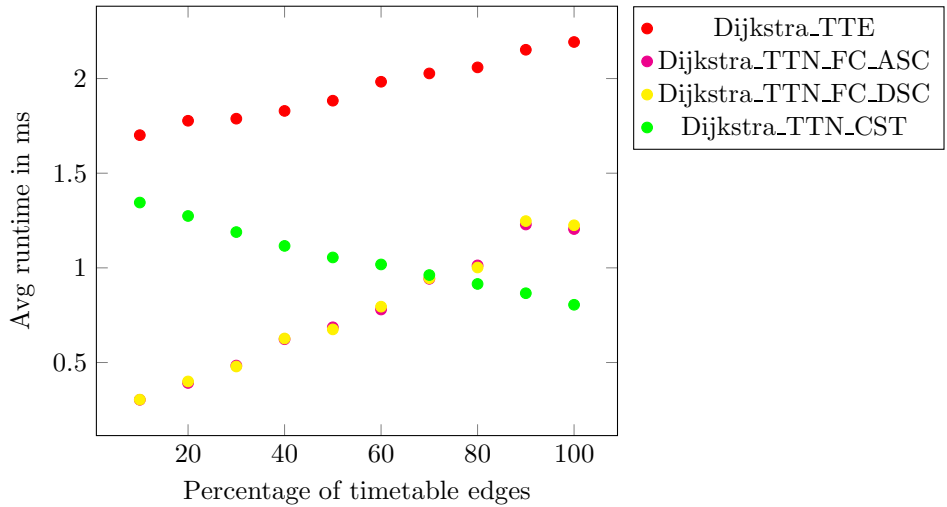
\begin{figure}
    \centering
    \begin{tikzpicture}
        \begin{axis}[
            title={Variation of percent of edges with timetable information},
            xlabel={Percentage of timetable edges},
            ylabel={Avg runtime in ms},
            legend pos=outer north east
        ]
        \addplot[
            only marks,
            color=red,
            mark=*,
            ]
            coordinates {
                (30, 1.788) (40, 1.829) (20, 1.777) (60, 1.983) (70, 2.027) (100, 2.193) (10, 1.701) (80, 2.059) (90, 2.152) (50, 1.883)
            };
        \addlegendentry{Dijkstra\_TTE}

        \addplot[
            only marks,
            color=magenta,
            mark=*,
            ]
            coordinates {
                (40, 0.623) (90, 1.229) (50, 0.686) (30, 0.484) (100, 1.205) (80, 1.013) (70, 0.942) (20, 0.392) (10, 0.303) (60, 0.78)
            };
        \addlegendentry{Dijkstra\_TTN\_FC\_ASC}
        
        \addplot[
            only marks,
            color=yellow,
            mark=*,
            ]
            coordinates {
                (20, 0.4) (30, 0.479) (50, 0.675) (40, 0.627) (80, 1.002) (90, 1.247) (10, 0.304) (100, 1.225) (60, 0.795) (70, 0.945)
            };
        \addlegendentry{Dijkstra\_TTN\_FC\_DSC}

        \addplot[
            only marks,
            color=green,
            mark=*,
            ]
            coordinates {
                (80, 0.915) (10, 1.345) (90, 0.866) (100, 0.805) (20, 1.274) (60, 1.018) (70, 0.962) (40, 1.116) (50, 1.055) (30, 1.189)
            };
        \addlegendentry{Dijkstra\_TTN\_CST}
        
        \end{axis}
    \end{tikzpicture}
    \caption{Variation of percent of edges with timetable information}
    \label{fig:runtime_vs_perc}
\end{figure}

To further investigate this hypothesis, we retained all walking connections between the nodes and varied the percentage of timetable edges. As a result, we observed the situation described in the Figure \ref{fig:runtime_vs_perc}.

On the graph, we noticed that TTN-FC outperforms other variations up to 70\% of timetable edges, after which TTN-CST performs better. We observed a similar trend in real cities examples, but the point at which one algorithm surpasses the other may differ. For example, in Berlin, TTN-CST performs better then TTN-FC on the original graph. This threshold may vary based on factors such as the total number of nodes, timetable structure, percentage of walking connections, and so on, which we did not vary in our synthetic graph experiment.

\section{Conclusion and Future Work}

 TTN is a good approach to solving multimodal pathfinding problems over high-density transport graphs with a large amount of timetable information. This algorithm takes a step forward in speeding up graph-based algorithms over the transport graph. The reason for it is the decreasing of asymptotic complexity of node evaluation from $ \mathcal{O} (k\times \log | C |) $ to $ \mathcal{O} (k + \log (k) + \log (| C |)) $. In the meantime, in the case of big cities with large numbers of nodes and outgoing edges, it is still better to use non-graph approaches, such as the Connection Scan Algorithm.

 In our future work, we plan to rewrite our solution in C++. We believe that this may lead to faster processing and better memory control. It will help that each Fractional Cascading tree be located in a compact memory region, consequently speeding up the search process. 

We plan to test our solution on the traditional road network problem in our future research. We want to see how well it performs in this context. Our approach could speed up the Time-Dependent Contraction Hierarchies with Compressed Database heuristic \cite{improving_time_dependent_ch, compressed_path_db}. Additionally, we're interested in comparing TTN with the method of splitting times into different buckets, which is currently widely used for dealing with the search over large time frames.

\subsection*{Acknowledgments}
Peter Stuckey was partially supported by the Australian Research Council OPTIMA ITTC IC200100009.



\bibliographystyle{plainnat}
\bibliography{timetable_nodes_for_public_transport}

\end{document}